\documentclass[aos,preprint]{imsart}

\RequirePackage{amsthm,amsmath,amsfonts,amssymb,bbm}
\RequirePackage[authoryear]{natbib} %% uncomment this for author-year bibliography
\RequirePackage[colorlinks,citecolor=blue,urlcolor=blue]{hyperref}
\RequirePackage{graphicx}

\startlocaldefs
\newtheorem{theorem}{Theorem}[section]
\newtheorem{lemma}[theorem]{Lemma}

\theoremstyle{remark}

\newtheorem{remark}[theorem]{Remark}
\newcommand{\norm}[1]{\left\lVert#1\right\rVert}
\newcommand{\Zt}{\tilde{Z}}
\def\E{\mathbb{E}} 
\def\P{\textsf{P}} 
\def\Cov{\textsf{Cov}} 
\def\Var{\textsf{Var}} 

\endlocaldefs
\begin{document}

\begin{frontmatter}
\title{UBSea: A Unified Community  Detection Framework}
%\title{A sample article title with some additional note\thanksref{t1}}
\runtitle{UBSea: A Unified Community  Detection Framework}
%\thankstext{T1}{A sample additional note to the title.}

\begin{aug}
%%%%%%%%%%%%%%%%%%%%%%%%%%%%%%%%%%%%%%%%%%%%%%
%%Only one address is permitted per author. %%
%%Only division, organization and e-mail is %%
%%included in the address.                  %%
%%Additional information can be included in %%
%%the Acknowledgments section if necessary. %%
%%%%%%%%%%%%%%%%%%%%%%%%%%%%%%%%%%%%%%%%%%%%%%
\author{\snm{XIANCHENG LIN AND HAO CHEN}}

%%%%%%%%%%%%%%%%%%%%%%%%%%%%%%%%%%%%%%%%%%%%%%
%% Addresses                                %%
%%%%%%%%%%%%%%%%%%%%%%%%%%%%%%%%%%%%%%%%%%%%%%
\address{
\normalsize{University of California, Davis}
}
\end{aug}
\begin{abstract}
  Detecting communities in networks and graphs is an important task across many disciplines such as statistics, social science and engineering.  There are generally three different kinds of mixing patterns for the case of two communities: assortative mixing, disassortative mixing and core-periphery structure.  Modularity optimization is a classical way for fitting network models with communities. However, it can only deal with assortative mixing and disassortative  mixing  when the mixing pattern is  known and fails to discover the core-periphery structure. In this paper, we extend modularity in a strategic way and propose a new framework based on \textbf{U}nified \textbf{B}igroups \textbf{S}tandadized \textbf{E}dge-count \textbf{A}nalysis (UBSea). It  can address all the formerly mentioned community mixing structures.  In addition, this new framework is able to automatically choose the mixing type to fit the networks. Simulation studies show that the new framework has superb performance in a wide range of settings under the stochastic block model  and the degree-corrected stochastic block model. We show that the new approach produces consistent estimate of the communities under a suitable signal-to-noise-ratio condition, for the case of a block model with two communities, for both undirected and directed networks.  The new method is illustrated through applications to several real-world datasets.
\end{abstract}


\end{frontmatter}
%%%%%%%%%%%%%%%%%%%%%%%%%%%%%%%%%%%%%%%%%%%%%%
%% Please use \tableofcontents for articles %%
%% with 50 pages and more                   %%
%%%%%%%%%%%%%%%%%%%%%%%%%%%%%%%%%%%%%%%%%%%%%%
%\tableofcontents
\section{Introduction}
%Detecting communities/clusters in networks and graphs is a fundamental task in statistics.
Network data analysis attracts attention  across many disciplines such as sociology, statistics, computer science and biology. One of the fundamental tasks in network data analysis is detecting communities/clusters. Many algorithms have been proposed for fitting network models with communities.
A standard framework for studying community detection in a statistical setting is the Stochastic Block Model (SBM) proposed in \cite{holland1983stochastic}. For a network with $N$ nodes, given its $N \times N$ adjacency matrix $\mathbf{A}$, this model assumes that the true node labels $\mathbf{c}=(c_{1},\cdots,c_{N}) \in\{1,\cdots,K\}^{N}$ are drawn independently from the multinomial distribution with proportion parameter $\boldsymbol{\pi}=(\pi_{1},\cdots,\pi_{K})$, where $\pi_{i}>0$ for all $i$, and $K$ is the number of communities, assumed known. Conditional on the labels assignment $\mathbf{c}$, the edge variables $A_{ij}$  independently follows Bernoulli distribution with
\begin{equation}
    \E[A_{ij}|\mathbf{c}]=P_{c_{i}c_{j}},
    \label{sto_prob}
\end{equation}
where $\mathbf{P}=[P_{ab}]$ is a $K\times K$ connectivity matrix. 
The task of community detection is then to infer the node labels $\mathbf{c}$ from the adjacency matrix $\mathbf{A}$, and often  involves estimating $\boldsymbol{\pi}$ and $\mathbf{P}$. In this paper, we will consider the  case where $K = 2$, i.e., two different communities.

There are many extensions of the block model, such as the Mixed Membership Stochastic Block Model in \cite{airoldi2008mixed}, the Weighted Stochastic Block Model in \cite{peixoto2018nonparametric}, to name but a few. One notable extension is the Degree-Corrected Stochastic Block Model (DCSBM) \citep{karrer2011stochastic}. It removes the unrealistic constraint of the same expected degree for all nodes within the same community in SBM by replacing (\ref{sto_prob}) with $\E[A_{ij}|\mathbf{c}]=\theta_{i}\theta_{j}P_{c_{i}c_{j}}$, where $\theta_{i}$'s are node degree parameters which satisfy an identifiability constraint such as $\mathbb{E}[\theta_{i}]=1$ or $\theta_{\max} = 1$. Researchers tend to replace the Bernoulli distribution for $A_{ij}$  by the Poisson distribution for the ease of technical derivations, which is in fact  a good approximation for a range of networks; see \cite{karrer2011stochastic}.

For the case of two communities, there are generally three types of mixing patterns under SBM for undirected networks. In the most commonly studied case, $P_{11}$ and $P_{22} $ are both larger than $P_{12}$, meaning that the probability for connections within communities are  larger than the probability between communities. This choice gives us the traditional community structure  called the assortative mixing \citep{newman2002assortative,newman2003mixing,newman2003structure}. To the opposite of the assortative mixing is the disassortative choice where $P_{11}$ and $P_{22}$ are both smaller than $P_{12}$, which means nodes are more likely to connect with nodes in the different communities. This disassortative mixing, commonly found in biological, technological networks and other fields, has received a modest amount of attention in the literature; see \cite{newman2007mixture,moore2011active}. There is, however, a third kind of mixing pattern in which $P_{11} > P_{12} > P_{22}$ (or $P_{22} > P_{12} > P_{11}$), which is called the core-periphery structure; see \cite{zhang2015identification}. Already in late 1990s, sociologists noticed that the core-periphery structure exhibited in a number of social networks; see \cite{borgatti2000models}. Without loss of generality, we can assume $P_{11}$ to be the largest of the three probabilities. Community one is then called the core and community two  the periphery. In the  core-periphery structure, edges are most likely to be within the core, least probable within the periphery, and modest number of edges between core and periphery.

Many algorithms have been proposed for fitting networks with community structures mentioned above, respectively. There are in general three categories of traditional methods: Spectral methods, likelihood-based methods and modularity-based methods. Spectral methods estimate the communities using the graph eigenvectors and are usually computationally fast; see \cite{von2007tutorial}. Spectral methods are known to suffer from inconsistency in sparse graphs \citep{krzakala2013spectral} and are often sensitive to outliers \citep{cai2015robust}. Profile likelihood maximization methods, for instance, \cite{wang2017likelihood}, are often  computationally intractable since they need to estimate a large number of parameters. One notable exception is the CPL method in \cite{amini2013pseudo}, which is efficient in solving large scale sparse networks. Methods based on modularity optimization have been found to produce excellent results. In general, modularity maximization targets at assortative mixing patterns; see \cite{newman2002assortative,newman2004finding,newman2006modularity}. Modularity minimization, on the contrary, discovers the disassortative mixing pattern; see \cite{newman2006finding}. However, modularity fails to discover core-periphery structure.  Researchers in modularity domain thereafter turn to likelihood methods to handle the core-periphery structure, making the modularity based community detection framework incomplete. Also, they most of the time only tested the performance under the simple stochastic block model; see \cite{zhang2015identification}. It is  worth pointing out that all these popular methods mentioned above are mainly designed for undirected networks and often require converting the directed graph to the undirected one, which in a sense, changes the graph information carried by the edges; see \cite{malliaros2013clustering} for a review. The most significant problem is that none of them provide a satisfying framework to include all the formerly mentioned mixing patterns simultaneously.  Moreover, people usually focus on one specific mixing pattern, mostly assortative mixing, to fit the network observation. Therefore, none of them can assess which mixing pattern is the most significant among the three mixing patterns.

The question remains, can we build a unified community detection framework to include all the following community structures and network types: assortative mixing, disassortative mixing, core-periphery structure, undirected networks and directed networks. The answer is surprisingly YES by revisiting the original definition of modularity. We extend modularity discussed in \cite{newman2006modularity} in a strategic way and propose a novel UBSea community detection framework which contains two closely related standardized clustering statistic $Z_{w}$ and $Z_{d}$, which will be maximized or minimized at the true clustering labels up to a permutation in high probability under proper conditions. The statistic $Z_{d}$, proposed in \cite{chen2017new}, was originally designed to overcome the difficulty in two-sample test  affected by the curse of dimensionality. Later, another edge-count statistic, the weighted edge-count statistic $Z_{w}$, was proposed in \cite{chen2018weighted}. In the two-sample test region, the statistic $Z_{w}$ focuses on the location alternative and the statistic $Z_{d}$ exhibits high power under a wide range of alternatives but is especially powerful in the scale alternative. Here in the community detection region,  we will see that, these two statistics play similar roles as in the two-sample test setting. The community detection performance based on the statistic $Z_{w}$ is similar to traditional community detection methods such as modularity methods or spectral clustering methods. The statistic $Z_{w}$ is mainly used to examine the assortative mixing and disassortative mixing patterns, while $Z_{d}$ mostly focus on the core-periphery structure. But they do have certain overlap and can both be powerful in a wide range of settings. As a result, when combined together, they form the foundation of our UBSea community detection framework. We prove that, the statistic $Z_{w}$ provides almost exact recovery of the communities, in the sense the misclustering fraction, up to a permutation, goes to zero, under a certain  condition of the signal-to-noise ratio (SNR). When the model is the symmetric stochastic block model with two communities, this condition on the connectivity matrix is equivalent to the one derived in  \cite{amini2013pseudo}. The statistic $Z_{d}$ provides almost exact recovery estimates of the communities, under a similar condition of SNR, for the case of two communities. We also prove exact recovery, in the sense that the number of misclustering nodes goes to zero, under slightly stronger conditions.

In order to build a complete framework, the last step is how to choose among $Z_{w}$, $Z_{d}$ and  other possibly related quantities in a statistical way. First, if the network follows the SBM, we found a $\gamma$-$\tau$ criterion, which originates from the above mentioned consistency results, can be utilized  to choose among those three mixing patterns. In a more general setting where the DCSBM better fits the network, we adopt the idea in \cite{wang2017likelihood}, in which they use the penalized likelihood to solve the problem of estimating the proper number of communities under SBM and DCSBM. We view the heterogeneity of the degree parameters $\theta$'s estimated by the given mixing pattern as the model complexity and thereafter develop a penalized likelihood criterion to choose among those three mixing patterns. We found that the penalized likelihood criterion to be more powerful than the $\gamma$-$\tau$ criterion in the DCSBM case.

Besides, our novel UBSea community detection framework is suitable for both undirected and directed networks. The problem of network community detection has mainly been considered and studied for the case of undirected networks. A huge number of diverse algorithms have been proposed for the undirected setting, involving contributions from the fields of computer science, statistics, social science, physics and biology \citep{aggarwal2010survey}.  Nevertheless, numerous graph data in several applications are by nature directed. The problem of community detection in directed networks is considered to be a more challenging task as compared to the undirected case; see the review \cite{malliaros2013clustering}. Traditional undirected methods convert the directed graphs to undirected graphs as the first step, which surely changes the information carried by the edges. While our framework can be applied to directed networks directly to fully use the direction information. Simulation results show that our UBSea framework is consistently powerful for both undirected and directed networks.

The rest of paper is organized as follows. We present the new method and consistency results in Section \ref{sec:method and theory}. The numerical performance of the methods is demonstrated in Section \ref{sec:Numerical result section}. Section \ref{sec:W-D framework section} states the unified community detection framework. We test the new framework on several real world datasets in Section \ref{sec:real data}. Section \ref{sec:conclusion and discussion} concludes with discussion. Section \ref{sec:proofs} provides the main steps of the proof for consistency results.

%We develop this framework under the simple SBM with two communities. While simulation results show that this framework also enjoys extremely good performance for  data generated from the Degree-Corrected Stochastic Block Model under a wide range of settings, which can be approximated by  the stochastic block model with multiple communities, making it promising to generalize this framework to SBM with multiple communities, typically for data generated from  Binary Tree Stochastic Block Model \citep{li2022hierarchical}. We leave the task of multiple communities detection under BTSBM as part of our future work.
\section{Proposed method and theory}
\label{sec:method and theory}
In this section, we first revisit the original definition of modularity and explain its intuition, as well as reasons for us to extend it. Then we introduce two statistics $R_{w}$ and $R_{d}$, which will be of central interest in our paper. We calculate their expectations and variances under the permutation null distribution. We show that the standardized versions of $R_{w}$ and $R_{d}$, called $Z_{w}$ and $Z_{d}$ can be used as community detection criteria. We determine the requirements of the growth rate on the connectivity matrix, in order to achieve almost exact recovery (weak consistency) and exact recovery (strong consistency), respectively, in the sense the misclustering fraction (weak consistency) or misclustering number (strong consistency) decay to 0 as node size increases, for the case of two communities. We state the results in this section and defer the consistency proofs in Section \ref{sec:proofs} and all other proofs in Supplementary Material \cite{Lin2023Supplement}.

\subsection{Revisit Modularity}
\label{revisit modularity}
There are many different forms of modularity. We here analyze the original form in \cite{newman2006modularity}:
\begin{equation}
\label{equ: modularity definition}
  \begin{aligned}
 Q&= \sum_{i,j}
 \left(A_{ij}-\frac{k_{i}k_{j}}{2m}\right) \frac{(s_{i}s_{j}+ 1)}{2}\\
 &=\text{(number of edges within communities)}  - \text{(expected number of such edges)}\\
  &=\left( R - \E R\right),
\end{aligned}  
\end{equation}
where $s_{i} = +1$ if vertex $i$ belongs to group 1 and $s_{i} = -1$ if it belongs to group 2, $k_{i},k_{j}$ are node degrees and $m$ is the total number of edges in the network. Let $R_{1}$ be the edges inside community 1 and $R_{2}$ be the edges inside community 2, then $R = R_{1} + R_{2}$ is the edges inside the same community. The choice of the ``expected number'' of edges is essentially equivalent to choosing a ``null model'' against which to compare the network. The null model behind modularity is close to the configuration model, where the edges are placed at random resulting in the network follows some pre-given degree distribution. Intuitively,  modularity measures the total deviation of edges inside each community from its expected number under the chosen ``null model''. Therefore, when there are two assortative communities, the edges inside them will both be larger than the expected number and modularity maximization is then able to capture this structure. On the contrary, when there are two disassortative communities, the edges inside them are both  smaller than the expected number, hence mudularity minimization is suitable in this case. However, in the core-periphery case, the deviation measurement  inside the core is positive while the deviation measurement inside the periphery is negative, hence they cancel each other out when summed up. Therefore, modularity maximization and minimization could not recognize the core-periphery structure. To sum up, the choice of the original modularity has the following drawbacks: First, it cannot deal with core-periphery structure as  explained above. Second, traditionally, people maximize modularity to find assortative communities. While minimizing modularity can find disassortative communities, it is then not clear when to maximize  or minimize it. Third, introducing the degree parameter $k_{i}$ then arising an issue about whether to use $k^{in}$ or $k^{out}$ in directed graphs; see \cite{leicht2008community} for a discussion. Last, researchers found that the normalizing term $\frac{1}{2m}$ might not work well under some common settings and they proposed to use a tuning parameter $\lambda$ to replace $\frac{1}{2m}$ in (\ref{equ: modularity definition}) \citep{reichardt2006statistical,fortunato2007resolution}, which abandoned the interpretability of the null model.  With these in mind, instead of sacrificing the interpretability of modularity to perform  modifications, our idea is to put modularity back to its general class, try different expectations, generalize $R$ in the following way:
\begin{itemize}
    \item We  adopt a different $\E A_{ij}$: The expectation under the permutation distribution, i.e., the null model is that there is only one community. 
    \item We generalize $R = R_{1} + R_{2}$  to be $\tilde{R} = aR_{1} + bR_{2}$, where $a$ and $b$ could be functions of the community sizes.
\end{itemize}
\subsection{Analysis of \texorpdfstring{$Z_{w}$}{Zw} and \texorpdfstring{$Z_{d}$}{Zd}}
We use $(G,E)$ to denote the  graph and the set of edges. When the graph is directed, $(i,j) \in E$ means that there is an edge pointing from node $i$ to node $j$. When the graph is undirected, $(i,j)\equiv (j,i)$, and $(i,j)\in E$ means there is an edge between nodes $i$ and $j$. We assume there is no self-loop in the graph. Let $\boldsymbol{x}$ be an $N$-length vector of 0's and 1's.
$$
R_{1}(\boldsymbol{x})=\sum_{(i,j)\in E}I_{x_{i}=x_{j}=1}, \quad R_{2}(\boldsymbol{x})=\sum_{(i,j)\in E}I_{x_{i}=x_{j}=0}.
$$

The vector $\boldsymbol{x}$ divides the observations into two communities: 1's (Community 1) and 0's (Community 2). Let $m_{\boldsymbol{x}}=\sum_{i=1}^{N}x_{i}$ be the number of vertices in the first community implied by $\boldsymbol{x}$, and $n_{\boldsymbol{x}}=N-m_{\boldsymbol{x}}$ be the number of vertices in the second community. Also, Let $\boldsymbol{x}^{*}$ represents the true community label, and there are $m$ 1's and $n$ 0's in $\boldsymbol{x}^{*}$, where $m$ and $n$ are the true community sizes. Let $\mathbf{P}=\begin{bmatrix}
P_{11}&P_{12}\\
P_{21}&P_{22}
\end{bmatrix}$ be the connectivity matrix, where $P_{ij}$ represents the probability of edge between community $i$ and community $j$. There are a number of reasonable choices for $a$ and $b$ in the definition of $\tilde{R}$ to generalize modularity. We here present two candidates named $R_{d}$ and $R_{w}$ to serve as an example to show the behavior of different choices through the analysis under stochastic block model. We will see that $R_{d}$ plays as the proof basis for other more complicated generalizations and $R_{w}$ is an example of taking consideration of the community sizes. To be more specific, define $ R_{w}(\boldsymbol{x})$ and $ R_{d}(\boldsymbol{x})$ and their normalized versions $Z_{w}(\boldsymbol{x})$ and $Z_{d}(\boldsymbol{x})$ to be the following terms:
\begin{alignat*}{2}
&Z_{w}(\boldsymbol{x}) = \frac{R_{w}(\boldsymbol{x})-\mu_{w}(\boldsymbol{x})}{\sigma_{w}(\boldsymbol{x})}, \quad &&R_{w}(\boldsymbol{x})=\frac{n_{\boldsymbol{x}}-1}{N-2} R_{1}(\boldsymbol{x})+\frac{m_{\boldsymbol{x}}-1}{N-2} R_{2}(\boldsymbol{x}), \\
&Z_{d}(\boldsymbol{x}) = \frac{R_{d}(\boldsymbol{x})-\mu_{d}(\boldsymbol{x})}{\sigma_{d}(\boldsymbol{x})}, \quad &&R_{d}(\boldsymbol{x})=R_{1}(\boldsymbol{x})-R_{2}(\boldsymbol{x}),
\end{alignat*}
where $\mu_{w}(\boldsymbol{x}),\sigma_{w}(\boldsymbol{x}),\mu_{d}(\boldsymbol{x}),\sigma_{d}(\boldsymbol{x})$ are normalizing terms. Their expressions can be obtained through similar treatments in \cite{chen2017new} and \cite{liu2022fast} and are provided below. 
\theoremstyle{plain} 
\begin{lemma}\label{thm2.1}
The expectation and variance of $R_{w}(\boldsymbol{x})$ and $R_{d}(\boldsymbol{x})$ on the  \textbf{directed} graph $(G,E)$ under permutation null are:
$$
\begin{aligned}
&\mu_{w}(\boldsymbol{x})\triangleq \E(R_{w}(\boldsymbol{x})) =\frac{(m_{\boldsymbol{x}}-1)(n_{\boldsymbol{x}}-1)}{(N-1)(N-2)}|G|,\\
&\sigma_{w}^{2}(\boldsymbol{x}) \triangleq \Var(R_{w}(\boldsymbol{x}))=\frac{m_{\boldsymbol{x}}n_{\boldsymbol{x}}(m_{\boldsymbol{x}}-1)(n_{\boldsymbol{x}}-1)}{N(N-1)(N-2)^2}\left(|G|+q_{1}-\frac{|G|^{2}}{N-1}+\frac{q_{2}}{N-3}\right),\\
&\mu_{d}(\boldsymbol{x})\triangleq \E(R_{d}(\boldsymbol{x})) = \frac{m_{\boldsymbol{x}}-n_{\boldsymbol{x}}}{N}|G|,\\
&\sigma_{d}^{2}(\boldsymbol{x}) \triangleq  \Var(R_{d}(\boldsymbol{x}))=\frac{m_{\boldsymbol{x}}n_{\boldsymbol{x}}}{N(N-1)}\left(|G|+q_{1}+|G|^{2}\frac{N-4}{N}-q_{2}\right),
\end{aligned}
$$
where $|G|,q_{1},q_{2}$ are quantities on the graph $(G,E)$:
\iffalse
c_{3}=c_{4}=\sum_{i=1}^{N}\sum_{j \in D_{i}^{s}} d_{j} \qquad\\
&c_{5}=\sum_{i=1}^{N}d_{i}(d_{i}-1) \qquad
c_{6}=\sum_{i=1}^{n}D_{i}(D_{i}-1) \qquad \qquad  c_{7}=c_{1}^{2}-\sum_{m=1}^{6}c_{m}\\
\fi
$$
\begin{aligned}
&|G|=\sum_{i=1}^{N}\sum_{j=1}^{N}\mathbbm{1}_{\{(i,j)\in E\}}, \qquad
q_{1}=\sum_{i=1}^{N} \sum_{j \in D_{i}^{s}} \mathbbm{1}_{\{(i, j) \in E\}},\\
&q_{2} = |G|^{2} - |G|-q_{1} - 2\sum_{i=1}^{N}\sum_{j \in D_{i}^{s}} k_{j}^{out} - \sum_{i=1}^{N}k_{i}^{out}(k_{i}^{out}-1) - \sum_{i=1}^{N}k_{i}^{in}(k_{i}^{in}  - 1).
\end{aligned}
$$
Here, $D_{i}^{s}$ is the set of vertices that point toward vertex $i$, $k_{i}^{in}$ is the cardinality of set $D_{i}^{s}$, or the in-degree of vertex $i$, $k_{i}^{out}$ is the out-degree of vertex $i$.
\label{E,var}
\end{lemma}

\theoremstyle{plain} 
\begin{lemma}\label{lemma2.2}
The expectation and variance of $R_{w}(\boldsymbol{x})$ and $R_{d}(\boldsymbol{x})$ on the  \textbf{undirected} graph $(\ddot{G},\ddot{E})$ under permutation null are:
$$
\begin{aligned}
&\mu_{w}(\boldsymbol{x})\triangleq \E(R_{w}(\boldsymbol{x})) =\frac{(m_{\boldsymbol{x}}-1)(n_{\boldsymbol{x}}-1)}{(N-1)(N-2)}|\ddot{G}|,\\
&\sigma_{w}^{2}(\boldsymbol{x}) \triangleq \Var(R_{w}(\boldsymbol{x}))=\frac{m_{\boldsymbol{x}}n_{\boldsymbol{x}}(m_{\boldsymbol{x}}-1)(n_{\boldsymbol{x}}-1)}{N(N-1)(N-2)^2}\left(|\ddot{G}|-\frac{|\ddot{G}|^{2}}{N-1}+\frac{\ddot{q}_{2}}{N-3}\right),\\
&\mu_{d}(\boldsymbol{x})\triangleq \E(R_{d}(\boldsymbol{x})) = \frac{m_{\boldsymbol{x}}-n_{\boldsymbol{x}}}{N}|\ddot{G}|,\\
&\sigma_{d}^{2}(\boldsymbol{x}) \triangleq  \Var(R_{d}(\boldsymbol{x}))=\frac{m_{\boldsymbol{x}}n_{\boldsymbol{x}}}{N(N-1)}\left(|\ddot{G}|+|\ddot{G}|^{2}\frac{N-4}{N}-\ddot{q}_{2}\right),
\end{aligned}
$$
where $|\ddot{G}|,\ddot{q}_{2}$ are quantities on the graph $(\ddot{G},\ddot{E})$:
\iffalse
c_{3}=c_{4}=\sum_{i=1}^{N}\sum_{j \in D_{i}^{s}} d_{j} \qquad\\
&c_{5}=\sum_{i=1}^{N}d_{i}(d_{i}-1) \qquad
c_{6}=\sum_{i=1}^{n}(D_{i}^{2}-D_{i}) \qquad \qquad  c_{7}=c_{1}^{2}-\sum_{m=1}^{6}c_{m}\\
\fi
$$
\begin{aligned}
&|\ddot{G}|=\sum_{i=1}^{N}\sum_{j=1}^{i}\mathbbm{1}_{\{(i,j)\in \ddot{E}\}}, \qquad
\ddot{q}_{2} = |\ddot{G}|^{2} - |\ddot{G}| - \sum_{i=1}^{N}k_{i}(k_{i}-1).
\end{aligned}
$$
Here, $k_{i}$ is the degree of vertex $i$.
\label{E,var,undirected}
\end{lemma}

\begin{figure}[b]
    \centering
    \includegraphics[scale=0.22]{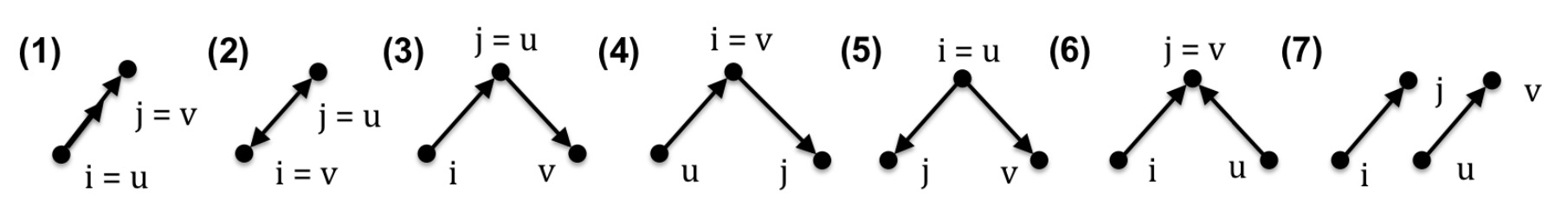}
    \caption{Seven possible configurations of two edges $(i,j)$,$(u,v)$ randomly chosen, with replacement, from a directed graph: (1) two edges degrenerate into one; (2) two opposite edges; (3)-(6) four different configurations with the two edges sharing one node; (7) two edges without any node sharing.}
    \label{cases}
\end{figure}

Lemma \ref{E,var} and Lemma \ref{E,var,undirected} can be proved by combinatorial analysis  similar in \cite{chen2017new} and \cite{liu2022fast}. The expectations are relatively simple due to the linearity of expectation. For the variances,  if the graph is directed, we have to figure out the number of possible configurations of pairs of edges plotted in Figure \ref{cases}. For undirected graphs, the treatment is slightly easier as the number of configurations to be considered is less.

\iffalse
We use $\E_{B;\mathbf{P},\mathbf{x^{*}}}(.)$ to denote the expectation under stochastic \textbf{B}lock model with connectivity matrix $\mathbf{P}$ and true labels assignment $\boldsymbol{x}^{*}$. Suppose $(G,E)$ is generated from such a model, then
\[
\E_{B;\mathbf{P},\mathbf{x^{*}}}[\mathbbm{1}_{(i,j) \in E}] = P_{c_{i}(\boldsymbol{x}^{*})c_{j}(\boldsymbol{x}^{*})}
\]
where $c_{i}(\boldsymbol{x}^{*}) = 1$ if $\boldsymbol{x}^{*}_{i} = 1$ otherwise $c_{i}(\boldsymbol{x}^{*}) = 2$.
We further define 
$$
\begin{aligned}
&R_{1}^{P}(\boldsymbol{x}) = \E_{B;\mathbf{P},\boldsymbol{x}^{*}}(R_{1}(\boldsymbol{x})),
\quad R_{2}^{P}(\boldsymbol{x}) = \E_{B;\mathbf{P},\boldsymbol{x}^{*}}(R_{2}(\boldsymbol{x})),
\quad |G_{B}| = \E_{B;\mathbf{P},\boldsymbol{x}^{*}}|G|,\\
&R_{d}^{P}(\boldsymbol{x}) = R_{1}^{P}(\boldsymbol{x}) - R_{2}^{P}(\boldsymbol{x}), \quad R_{w}^{P}(\boldsymbol{x})  = \frac{n_{\boldsymbol{x}}-1}{N-2}R_{1}^{P}(\boldsymbol{x}) + \frac{m_{\boldsymbol{x}}-1}{N-2}R_{2}^{P}(\boldsymbol{x}),\\
&\mu_{d}^{P}(\boldsymbol{x}) = \frac{m_{\boldsymbol{x}} - n_{\boldsymbol{x}}}{N}|G_{B}|, \qquad \qquad \mu_{w}^{P}(\boldsymbol{x}) = \frac{(m_{\boldsymbol{x}} - 1)(n_{\boldsymbol{x}}-1)}{(N-1)(N-2)}|G_{B}|,\\
&Z_{d}^{P}(\boldsymbol{x}) = \frac{R_{d}^{P}(\boldsymbol{x})-\mu_{d}^{P}(\boldsymbol{x})}{\sigma_{d}(\boldsymbol{x})}, \qquad
Z_{w}^{P}(\boldsymbol{x}) = \frac{R_{w}^{P}(\boldsymbol{x})-\mu_{w}^{P}(\boldsymbol{x})}{\sigma_{w}(\boldsymbol{x})}.
\end{aligned}
$$
\fi
Following the treatment in \cite{chen2023new}, we can define $\E_{2}(\cdot)$ to be the expectation under the two communities SBM with the connectivity matrix $\mathbf{P}$. We let
\[
R_{d,c}^{P}(\boldsymbol{x}) = \E_{2}(R_{d}(\boldsymbol{x})-\mu_{d}(\boldsymbol{x})), \quad R_{w,c}^{P}(\boldsymbol{x}) = \E_{2}(R_{w}(\boldsymbol{x})-\mu_{w}(\boldsymbol{x})).
\]

For directed graphs, given the label assignment $\boldsymbol{x}$ and true label $\boldsymbol{x}^{*}$, suppose there are $\Delta_{1}\in[0,m]$ observations in community 1 placed into community 2, and $\Delta_{2}\in[0,n]$ observations in community 2 placed into community 1. The associated sizes of the two communities of assignment $\boldsymbol{x}$ are 
\[
m_{\boldsymbol{x}}=m-\Delta_{1}+\Delta_{2}, \quad n_{\boldsymbol{x}}=n-\Delta_{2}+\Delta_{1}.
\]
Then,
$$
\begin{aligned}
&\E_{2}(R_{1}(\boldsymbol{x}))=(m-\Delta_{1})(m-\Delta_{1}-1)P_{11}+(m-\Delta_{1})\Delta_{2}(P_{12}+P_{21})+\Delta_{2}(\Delta_{2}-1)P_{22},\\
&\E_{2}(R_{2}(\boldsymbol{x}))=(n-\Delta_{2})(n-\Delta_{2}-1)P_{22}+(n-\Delta_{2})\Delta_{1}(P_{12}+P_{21})+\Delta_{1}(\Delta_{1}-1)P_{11},\\
&\E_{2}(\mu_{d}(\boldsymbol{x}))=\frac{(m-n)-2(\Delta_{1}-\Delta_{2})}{N}(m(m-1)P_{11}+mn(P_{12}+P_{21})+n(n-1)P_{22}),\\
&\E_{2}(\mu_{w}(\boldsymbol{x}))=\frac{(m-\Delta_{1}+\Delta_{2}-1)(n-\Delta_{2}+\Delta_{1}-1)}{(N-1)(N-2)}(m(m-1)P_{11}+mn(P_{12}+P_{21})\\
&\qquad \qquad \qquad +n(n-1)P_{22}),\\
&\E_{2}(R_{w}(\boldsymbol{x})-\mu_{w}(\boldsymbol{x}))\\
&=\frac{1}{N-2}\{P_{11}((n-\Delta_{2}+\Delta_{1}-1)(m-\Delta_{1})(m-\Delta_{1}-1)\\
&\quad +(m-\Delta_{1}+\Delta_{2}-1)\Delta_{1}(\Delta_{1}-1))+P_{22}((n-\Delta_{2}+\Delta_{1}-1)\Delta_{2}(\Delta_{2}-1)\\
&\quad+(m-\Delta_{1}+\Delta_{2}-1)(n-\Delta_{2})(n-\Delta_{2}-1))\\
&\quad+(P_{12}+P_{21})((m-\Delta_{1})\Delta_{2}(n-\Delta_{2}+\Delta_{1}-1)+(n-\Delta_{2})\Delta_{1}(m-\Delta_{1}+\Delta_{2}-1)\\
&\quad-\frac{(m-\Delta_{1}+\Delta_{2}-1)(n-\Delta_{2}+\Delta_{1}-1)}{(N-1)}(m(m-1)P_{11}+mn(P_{12}+P_{21})\\
&\quad +n(n-1)P_{22}))\}\\
&=\frac{mn(m-1)(n-1)}{(N-1)(N-2)}(P_{11}+P_{22}-P_{12}-P_{21})\\
    &\quad \times \left(1+\frac{\Delta_{1}^{2}}{m(m-1)}+\frac{\Delta_{2}^{2}}{n(n-1)}-\frac{(2m-1)\Delta_{1}}{m(m-1)}-\frac{(2n-1)\Delta_{2}}{n(n-1)}+\frac{2\Delta_{1}\Delta_{2}}{mn}\right),
\end{aligned}
$$

$$
\begin{aligned}
&\E_{2}(R_{d}(\boldsymbol{x})-\mu_{d}(\boldsymbol{x}))\\
&=P_{11}((m-\Delta_{1})(m-\Delta_{1}-1)-\Delta_{1}(\Delta_{1}-1))\\
&\quad +P_{22}(\Delta_{2}(\Delta_{2}-1)-(n-\Delta_{2})(n-\Delta_{2}-1)) \\
&\quad  +(P_{12}+P_{21})((m-\Delta_{1})\Delta_{2}-(n-\Delta_{2})\Delta_{1})\\
&\quad -\frac{(m-n)-2(\Delta_{1}-\Delta_{2})}{N}(m(m-1)P_{11}+mn(P_{12}+P_{21})+n(n-1)P_{22})\\
&=\frac{mn}{m+n}(2(m-1)P_{11}-2(n-1)P_{22}-(m-n)(P_{12}+P_{21}))\left(1-\frac{\Delta_{1}}{m}-\frac{\Delta_{2}}{n}\right).
\end{aligned}
$$

The expressions of them for undirected graphs can be derived almost identically and are omitted here.
Now, we are ready to introduce the optimization results for $R_{w,c}^{P}(\boldsymbol{x})$ and $R_{d,c}^{P}(\boldsymbol{x})$.
\theoremstyle{plain} 
\begin{theorem}\label{thm2.3}
Let $\boldsymbol{x}^{*}$, $m$, $n$ denote the true community label and the corresponding number of vertices in community 1's and community 0's, respectively:
\begin{itemize}
    \item When $2(m-1)P_{11}-2(n-1)P_{22}-(m-n)(P_{12}+P_{21})\neq 0$, $R_{d,c}^{P}(\boldsymbol{x})$ and $R_{d,c}^{P}(\boldsymbol{x})/\sigma_{d}(\boldsymbol{x})$ are maximized at $\boldsymbol{x}=\boldsymbol{x}^{*}$ or $\boldsymbol{x}=1-\boldsymbol{x}^{*}$.
    \item When $P_{11}+P_{22}-P_{12}-P_{21}>0$, $R_{w,c}^{P}(\boldsymbol{x})$ and $R_{w,c}^{P}(\boldsymbol{x})/\sigma_{w}(\boldsymbol{x})$ are maximized at $\boldsymbol{x}=\boldsymbol{x}^{*}$ or $\boldsymbol{x}=1-\boldsymbol{x}^{*}$.
    \item When $P_{11}+P_{22}-P_{12}-P_{21}<0$, $R_{w,c}^{P}(\boldsymbol{x})$ and $R_{w,c}^{P}(\boldsymbol{x})/\sigma_{w}(\boldsymbol{x})$ are minimized at $\boldsymbol{x}=\boldsymbol{x}^{*}$ or $\boldsymbol{x}=1-\boldsymbol{x}^{*}$.
\end{itemize}
\theoremstyle{plain}
\begin{remark}
\label{remark:2.3}
The condition for $R_{d,c}^{P}(\boldsymbol{x})$ can be approximated by $\tfrac{m}{n}=\tfrac{P_{12}+P_{21}-2P_{22}}{P_{12}+P_{21}-2P_{11}}$. When $m=n$, and the connectivity matrix is symmetric, this is to say $P_{11}=P_{22}$.
\end{remark}
\end{theorem}

The proof of Theorem \ref{thm2.3} follows directly from Theorem 1 in \cite{chen2023new} and is provided in Supplementary Material \citep{Lin2023Supplement}.  For simplicity, in the following we denote  $R_{d,c}^{P}(\boldsymbol{x})/\sigma_{d}(\boldsymbol{x})$, $R_{w,c}^{P}(\boldsymbol{x})/\sigma_{w}(\boldsymbol{x})$ as $Z_{d}^{P}(\boldsymbol{x})$ and $Z_{w}^{P}(\boldsymbol{x})$, respectively. The optimization properties of $Z_{w}^{P}(\boldsymbol{x})$ and $Z_{d}^{P}(\boldsymbol{x})$ are crucial in establishing the consistency results for $Z_{w}(\boldsymbol{x})$ and $Z_{d}(\boldsymbol{x})$ in Section \ref{sec:consistency results}. Additionally, it is noteworthy that  these conditions, derived from $Z_{d}^{P}(\boldsymbol{x})$ and $Z_{w}^{P}(\boldsymbol{x})$, are often to be effective for $Z_{d}(\boldsymbol{x})$ and $Z_{w}(\boldsymbol{x})$  in practice.

\subsection{Consistency Result}
\label{sec:consistency results}
Denote $\mathbb{\chi_{N}}$ as the collection of $\{0,1\}^{N}$. Denote the true label as $\boldsymbol{x}^{*} \in \mathbb{\chi_{N}}$ with the community sizes $m \geq 2$ and $n \geq 2$, $N=m+n$ is the total node size. For a label $\boldsymbol{x}\in\mathbb{\chi_{N}}$ with $\Delta_{1}\in[0,m]$ observations in community 1 placed into community 2, and $\Delta_{2}\in[0,n]$ observations in community 2 placed into community 1, then $m_{\boldsymbol{x}}=m-\Delta_{1}+\Delta_{2}$, $n_{\boldsymbol{x}}=n-\Delta_{2}+\Delta_{1}$. Let $\delta_{1}=\frac{\Delta_{1}}{m} \in [0,1]$, $\delta_{2}=\frac{\Delta_{2}}{n} \in [0,1]$ be the corresponding misclustering proportion. Besides, we define $a_{n}=o(b_{n})$  or $a_{n} \ll b_{n}$ as $a_{n}$ is dominated by $b_{n}$ asymptotically i.e. $\lim_{n \rightarrow \infty} \frac{a_{n}}{b_{n}}=0$, $a_{n}=O(b_{n})$ or $a_{n} \asymp b_{n}$ as $a_{n}$ is bounded both above and below by $b_{n}$ asymptotically, and $a_{n} \lesssim b_{n}$ as $a_{n}$ is bounded above by $b_{n}$(up to constant factor) asymptotically.

We analyze the consistency property under a block model as the number of nodes in the graph $G$ grows.  We condition on the community labels $\{c_{i}\}$, that is, we treat them as deterministic unknown parameters. Recall that for the block model, we assume all the entries in the adjacency matrix are drawn independently from Bernoulli distribution, that is, 
$$
A_{ij}\sim Ber(P_{c_{i}c_{j}}) \quad \text{for all $i,j$}.
$$ 
where $c_{i} \in \{1,2\}$ corresponding to community 1 and community 2. Here in the proof, we assume that diagonal entries of the adjacency matrix are also drawn randomly (i.e., we allow for self-loops as valid with-community edges). This is convenient in the analysis with minor effect on the results. Our approach is to prove  consistency result for the SBM model,  with the connectivity matrix of the form:
$$
    \mathbf{P}=
    \begin{pmatrix}
    P_{11}&P_{12}\\P_{21}&P_{22}
    \end{pmatrix}.
$$
Here, all $P_{ij}$'s depend on $N$ and can change with $N$ at different rates. Note that we don't have additional restrictions on the connectivity matrix. This is different from the literature where they usually consider the Symmetric Stochastic Block Model (SSBM), where the diagonal and off-diagonal elements of the connectivity matrix are the same, such as \cite{amini2013pseudo}. We will see that, our consistency requirement is more generic and is equivalent to the condition on the connectivity matrix derived in \cite{amini2013pseudo} when the model is SSBM.

The notion of consistency of community detection as the number of nodes grows was introduced in \cite{bickel2009nonparametric}. They defined a community detection criterion $Q$ to be consistent if the node labels obtained by maximizing the criterion, $\hat{\mathbf{c}}=\arg \max _{\boldsymbol{x}} Q(\boldsymbol{x})$, satisfy 
\begin{equation}
     P[\hat{\mathbf{c}}=\mathbf{c}] \rightarrow 1 \quad \text { as } N \rightarrow \infty.
     \label{strong_consistency}
\end{equation}

Strictly speaking, this definition suffers from an identifiability issue. Because we can always do communities permutation. Thus, a better way to define consistency is to replace the equality $\mathbf{\hat{c}} = \mathbf{c}$ with the requirement that $\mathbf{\hat{c}}$ and $\mathbf{c}$ belong to the same equivalence class of label permutations. For simplicity of notation, we still write $\mathbf{\hat{c}} = \mathbf{c}$ below.

The notion of consistency in \cite{bickel2009nonparametric} is very strong, since it requires asymptotically no errors. Instead, we can also define the following notion of consistency,
\begin{equation}
    \forall \varepsilon>0 \quad \mathbb{P}\left[\left(\frac{1}{N} \sum_{i=1}^{N} \mathbb{I}\left(\hat{c}_{i} \neq c_{i}\right)\right)<\varepsilon\right] \rightarrow 1 \quad \text { as } N \rightarrow \infty.
    \label{weak consistency}
\end{equation}
where equality is also interpreted to mean labels in the same equivalence class with respect to label permutations.

Consistency pattern (\ref{weak consistency}) was called weak consistency  in \cite{zhao2012consistency}, in the sense that the misclustering fraction goes to zero as the graph node size increases. Consistency pattern (\ref{strong_consistency}) is the strong consistency used in \cite{karrer2011stochastic}, which means the clustering label is the same as the true label with high probability as the graph node size increases. By defining  $\gamma_{N}^{2}=\frac{(2\pi_{1}P_{11}-2\pi_{2}P_{22}-(\pi_{1}-\pi_{2})(P_{12}+P_{21}))^{2}}{\max(P)}$, $\tau_{N}^{2}=\frac{(P_{11}+P_{22}-P_{12}-P_{21})^{2}}{\max(P)}$, we will present both weak and strong consistency results for our method.
\theoremstyle{plain}

\theoremstyle{plain}
\begin{theorem}[Weak Consistency]
Under the SBM, 
\label{thm_weak_consistency}
\begin{itemize}
    \item If $N\gamma_{N}^{2} \rightarrow \infty$, then $\hat{\mathbf{c}}=\arg \max _{\boldsymbol{x}}Z_{d}(\boldsymbol{x})$ is weakly consistent under the assumption that there exists a sequence $\alpha_{N}$ with $\alpha_{N} \rightarrow 0$ and $\alpha_{N}N\gamma_{N}^{2} \rightarrow \infty$ such that \begin{equation}
        N e^{-\alpha_{N}N\gamma_{N}^{2}} < m_{\boldsymbol{x}},n_{\boldsymbol{x}} <N (1-e^{-\alpha_{N}N\gamma_{N}^{2}}),
        \label{assumption_1}
    \end{equation} 
    where $m_{\boldsymbol{x}}, n_{\boldsymbol{x}}$ are the corresponding group sizes of $\arg\max_{\boldsymbol{x}}Z_{d}(\boldsymbol{x})$. 
    \item If $N\tau_{N}^{2} \rightarrow \infty$ and $P_{11} + P_{22} - P_{12}-P_{21} > 0$, then $\hat{\mathbf{c}}=\arg \max _{\boldsymbol{x}}Z_{w}(\boldsymbol{x})$ is weakly consistent under the assumption that there exists a sequence $\beta_{N}$ with $\beta_{N} \rightarrow 0$ and $\beta_{N}N\tau_{N}^{2} \rightarrow \infty$  such that 
    \begin{equation}
       N e^{-\beta_{N}N\tau_{N}^{2}} < m_{\boldsymbol{x}},n_{\boldsymbol{x}} < N(1- e^{-\beta_{N}N\tau_{N}^{2}}),
        \label{assumption_2}
    \end{equation}
    where $ m_{\boldsymbol{x}}, n_{\boldsymbol{x}}$ are the corresponding group sizes of $\arg\max_{\boldsymbol{x}}Z_{w}(\boldsymbol{x})$.
    \item If $N\tau_{N}^{2} \rightarrow \infty$ and $P_{11} + P_{22} - P_{12}-P_{21} < 0$, then $\hat{\mathbf{c}}=\arg \min _{\boldsymbol{x}}Z_{w}(\boldsymbol{x})$ is weakly consistent under the assumption that there exists a sequence $\omega_{N}$ with $\omega_{N} \rightarrow 0$ and $\omega_{N}N\tau_{N}^{2} \rightarrow \infty$  such that 
    \begin{equation}
       N e^{-\omega_{N}N\tau_{N}^{2}} < m_{\boldsymbol{x}},n_{\boldsymbol{x}} < N(1- e^{-\omega_{N}N\tau_{N}^{2}}),
        \label{assumption_3}
    \end{equation}
where $ m_{\boldsymbol{x}}, n_{\boldsymbol{x}}$ are the corresponding group sizes of $\arg\min_{\boldsymbol{x}}Z_{w}(\boldsymbol{x})$.
\end{itemize}
\end{theorem}
\theoremstyle{plain}
\begin{theorem}[Strong Consistency]
\label{thm_strong_consistency}
Under the SBM, 
\begin{itemize}
    \item If $\frac{N\gamma_{N}^{2}}{\log(N)}\rightarrow \infty$, then $\hat{\mathbf{c}}=\arg \max _{\boldsymbol{x}}Z_{d}(\boldsymbol{x})$ is strongly consistent.
    \vspace{5pt}
    \item If $\frac{N\tau_{N}^{2}}{\log(N)}\rightarrow \infty$ and  $P_{11} +P_{22} -P_{12}-P_{21} > 0$, then $\hat{\mathbf{c}}=\arg \max _{\boldsymbol{x}}Z_{w}(\boldsymbol{x})$ is strongly consistent.
    \vspace{5pt}
    \item If $\frac{N\tau_{N}^{2}}{\log(N)}\rightarrow \infty$ and $P_{11}+P_{22} - P_{12}-P_{21} <0 $, then $\hat{\mathbf{c}}=\arg \min _{\boldsymbol{x}}Z_{w}(\boldsymbol{x})$ is strongly consistent.
\end{itemize}
\end{theorem}

\theoremstyle{remark}
\begin{remark}
In \cite{amini2013pseudo}, they consider the SSBM with $P_{11} = P_{22}$ and $P_{12} = P_{21}$.  They obtain the weak consistency condition $N\tilde{\tau}^{2} = N\frac{(P_{11} - P_{12})^{2}}{P_{11} + P_{12}} \rightarrow \infty$, which is equivalent with the weak consistency condition we derived here because $\frac{1}{2}\tau^{2}< \tilde{\tau}^{2} < \tau^{2}$ in this case.
\end{remark}
The proof of Theorem \ref{thm_weak_consistency} and Theorem \ref{thm_strong_consistency} follow the spirit in \cite{zhao2012consistency} and are provided in Section \ref{sec:proofs}. %We have obtained the optimization properties for  $Z_{w}^{P}$ and $Z_{d}^{P}$ in Theorem \ref{thm2.3}. Moreover, we can actually calculate the exact value of $Z_{d}^{P}(\boldsymbol{x}^{*})-Z_{d}^{P}(\boldsymbol{x})$ and $Z_{w}^{P}(\boldsymbol{x}^{*})-Z_{w}^{P}(\boldsymbol{x})$. Therefore, the optimization properties of $Z_{w}(\boldsymbol{x})$ and $Z_{d}(\boldsymbol{x})$ are analyzed by comparing the deviations $Z_{d}(\boldsymbol{x}) - Z_{d}^{P}(\boldsymbol{x})$, $Z_{w}(\boldsymbol{x}) - Z_{w}^{P}(\boldsymbol{x})$ with $Z_{d}^{P}(\boldsymbol{x}^{*})-Z_{d}^{P}(\boldsymbol{x})$, $Z_{w}^{P}(\boldsymbol{x}^{*})-Z_{w}^{P}(\boldsymbol{x})$.  
For the strong consistency, we don't have any additional assumptions except for $\gamma_{N}, \tau_{N}, N$. This can be seen by choosing $\alpha_{N} = \frac{\log(N)}{N\gamma_{N}^{2}}, \beta_{N} = \frac{\log(N)}{N\tau_{N}^{2}}$.

\section{Performance exploration}
\label{sec:Numerical result section}
In this section, we explore the performance of $Z_{w}$ and $Z_{d}$ on a variety of synthetic datasets, including assortative mixing, disassortative mixing, core-periphery structure,  directed graphs, undirected graphs, balanced communities, unbalanced communities, dense graphs and sparse graphs.

We use a greedy method for dividing networks in two by $Z_{w}^{max}$ (or $Z_{w}^{min},Z_{d}$): Start from a guess of $\boldsymbol{x}$ (could be randomly generated), we find among all single changes to $\boldsymbol{x}$ that is defined as changing one element in $\boldsymbol{x}$ from $s$ to $1-s$, and choose the change that will give the biggest increase (or decrease if using $Z_{w}^{min}$) in the corresponding $Z_{w}$ (or $Z_{d}$). We make such moves repeatedly until there is no increase (or decrease if using $Z_{w}^{min}$) available. We pick many random initial guesses of $\boldsymbol{x}$ and pick the result with maximum (or minimum if using $Z_{w}^{min}$) $Z_{w}$ (or $Z_{d}$). Running many initial guesses sometimes can be  time consuming when the number of vertices are pretty large. We suggest using the label assignment from CPL \citep{amini2013pseudo} as the initial guess to speed up the algorithm. In this section, we test the performance of $Z_{w}^{max},Z_{w}^{min},Z_{d}$ separately. We will discuss the method to automatically choose among these three statistics in Section \ref{sec:W-D framework section}.

We provide experiment results on  synthetic data generated from DCSBM, which contains much more flexibility than SBM. We specify the degree heterogeneity parameter $\theta_{i}$ to follow the pareto distribution in this section. We also try other distributions for $\theta_{i}$ and the results are in Supplementary Material \citep{Lin2023Supplement}. For each node $i \in [n]$,  $\theta_{i}$ is sampled independently from a Pareto($\alpha, \beta$) distribution with the density function $f(x \mid \alpha, \beta)=\frac{\alpha \beta^{\alpha}}{x^{\alpha+1}} \mathbf{1}_{\{x \geq \beta\}}$, where $\alpha$ and $\beta$ are called shape and scale parameters, respectively. The shape and scale parameters are adjusted so that the expectation of each $\theta_{i}$ is fixed at 1 to satisfy the identifiability constraint. Given the degree heterogeneity parameters $\{\theta_{i}\}$ and the connectivity matrix $\mathbf{P}$, a graph is generated from DCSBM, with the edge probability points from node $i \in C_{a}$ to node $j \in C_{b}$ being $\min(1, \theta_{i}\theta_{j}P_{ab})$. Note that when all the $\theta_{i} = 1$, the model becomes the standard SBM.

For comparison, we also apply several algorithms which are reported to have state-of-the-art empirical performance on DCSBM in the existing literature. Note that these methods represent three different categories of community detection methods named modularity-based methods, spectral clustering methods, and likelihood-based methods. Other methods in the same category are likely to have similar performance, hence these methods are representative for comparing out framework with the state-of-the-art literature:
\begin{enumerate}
    \item Convexified Modularity Maximization (CMM) algorithm in \cite{chen2018convexified} is based on convexification of the modularity maximization formulation and a doubly-weighted $l_{1}$ norm k-modoids clustering procedure.
    
    \item Modularity Minimization (MMin) in \cite{newman2006finding} is  a modularity minimization algorithm, targets at finding disassortative communities.
    
    \item The regularized spectral clustering (Spectral) in \cite{qin2013regularized}.
    
    \item The conditional pseudo-likelihood (CPL) maximization algorithm in \cite{amini2013pseudo} maximizes the conditional pseudo-likelihood via EM algorithm, starting with an initial partition provided by regularized spectral clustering.

    %\item The Zhang-Newman (ZN) algorithm in \cite{zhang2015identification} fits a generative model of core-periphery structure using a combination of an expectation-maximization algorithm and a belief propagation algorithm.
\end{enumerate}
The synthetic experiments were performed with different connectivity matrix $\mathbf{P}$'s, representing assortative mixing, disassortative mixing and core-periphery structure. We consider both directed graphs and undirected graphs, balanced design and unbalanced design, dense and sparse graphs in our simulation studies.  Note that as the shape parameter of the pareto distribution $\alpha$ increases, the degree parameters $\{\theta_{i}\}$ become less heterogeneous, hence the DCSBM becomes closer to the standard SBM.

Based on Theorem \ref{thm2.3}, \ref{thm_weak_consistency}, \ref{thm_strong_consistency}, we can now test the performances of $Z_{w}^{max}, Z_{w}^{min},Z_{d}$ separately by generating corresponding assortative mixing, disassortative mixing and core-periphery structure. Notice that $Z_{w}$ and $Z_{d}$ are not mutually exclusive, which means there are plenty of cases where they are both suitable for discovering the underlying communities, theoretically. Nevertheless, according to practical performance, the statistic $Z_{w}$ is usually preferred when the mixing type is assortative mixing or disassortative mixing, and $Z_{d}$ is useful for the core-periphery structure.

Figures \ref{Figure:equal size,pareto,dense, assorataive mixing}, \ref{Figure:equal size,pareto,dense, disassortative mixing}, \ref{Figure:equal size,pareto,dense, core-periphery} show that the misclassification rate of $Z_{w}^{max},Z_{w}^{min},Z_{d}$ all decrease as the degree parameters $\{\theta_{i}\}$ become less heterogeneous (larger value of the shape parameter). In all plots, each point represents the average of 50 independent runs. First, in the assortative mixing graph case, as indicated in Figure \ref{Figure:equal size,pareto,dense, assorataive mixing}, $Z_{w}^{max}$ achieves the highest power for both undirected and directed graphs. CMM, a modularity based method, which is designed for undirected graphs, also has good performance under directed graphs. Spectral methods and CPL are able to be aligned with $Z_{w}^{max}$ and CMM methods for undirected case, but will have a certain gap with the other two methods  when applied to directed graphs. MMin method is not suitable for assortative mixing case, hence being the worst here.
\begin{figure}[b]
\centering
\begin{minipage}[t]{0.48\textwidth}
\centering
\includegraphics[width=6cm, height = 5cm]{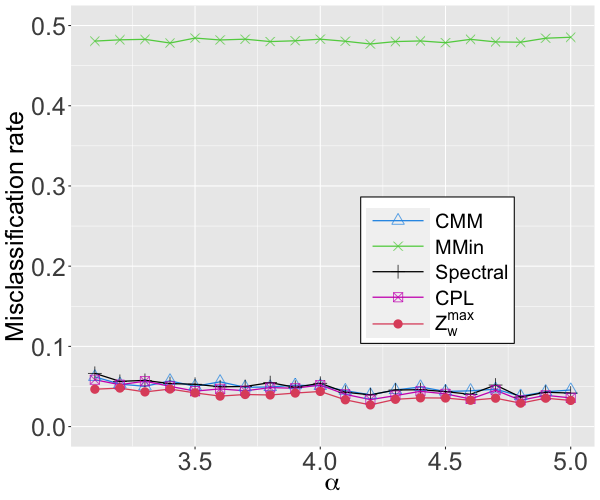}
\end{minipage}
\begin{minipage}[t]{0.48\textwidth}
\centering
\includegraphics[width=6cm, height = 5cm]{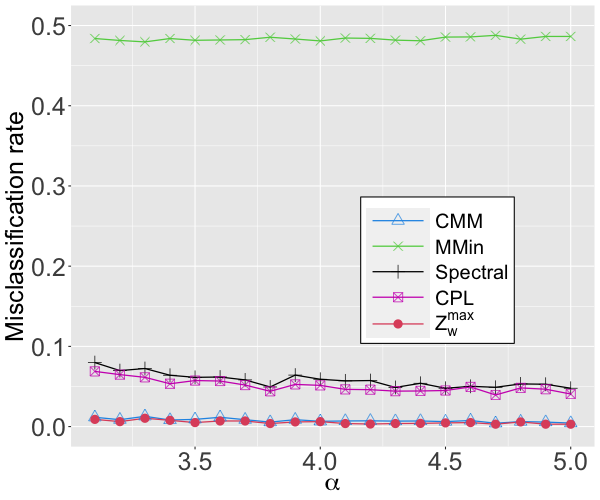}
\end{minipage}
\caption{Misclassification rate versus shape parameter of pareto $\theta$ for $N=100$ and 2 equal-sized communities. $\mathbf{P}= [0.5,0.3;0.3,0.5]$.  Left panel: undirected graph. Right panel: directed graph.}
\label{Figure:equal size,pareto,dense, assorataive mixing}
\end{figure}

For the disassortative mixing pattern, Figure \ref{Figure:equal size,pareto,dense, disassortative mixing} shows that  $Z_{w}^{min}$ has slightly better accuracy than CPL and Spectral method for undirected case. MMin performs slightly better than $Z_{w}^{min}$ in the undirected case and they are almost identical in the directed case. CMM completely loses its power for both undirected and directed graphs for the case of disassortative mixing. This is not surprising because CMM is essentially a modularity maximization algorithm. MMin and $Z_{w}^{min}$ outperform all other methods for directed graphs. Generally speaking, MMin, $Z_{w}^{min}$, Spectral and CPL all have satisfying performance for disassortative mixing case, while MMin and $Z_{w}^{min}$ are preferred for directed graphs.

\begin{figure}[htbp]
\centering
\begin{minipage}[t]{0.48\textwidth}
\centering
\includegraphics[width=6cm, height = 5cm]{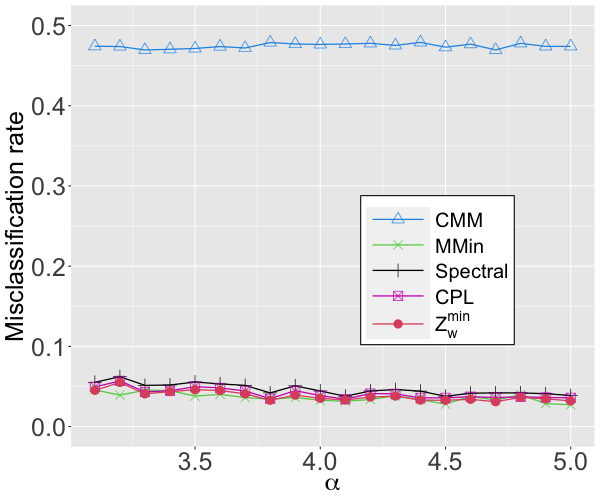}
\end{minipage}
\begin{minipage}[t]{0.48\textwidth}
\centering
\includegraphics[width=6cm, height = 5cm]{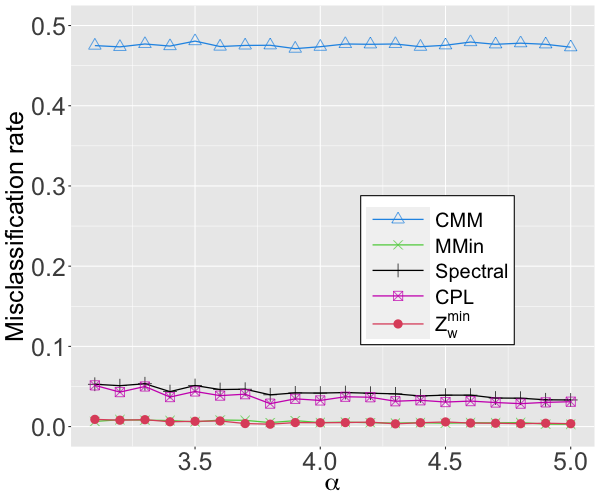}
\end{minipage}
\caption{Misclassification rate versus shape parameter of pareto $\theta$ for $N=100$ and 2 equal-sized communities. $\mathbf{P}=[0.3,0.5;0.5,0.3]$.  Left panel: undirected graph. Right panel:directed graph.}
\label{Figure:equal size,pareto,dense, disassortative mixing} 
\end{figure}

In the case of the core-periphery structure, all other methods lose power hence there is a significant power gap between $Z_{d}$ and all other methods. The misclassification rate for CMM, CPL and Spectral methods for both undirected and directed graphs are near 50\%, which means they are not much better than random guess. The reason behind the failure of these methods are that the periphery part are loosely connected inside them, hence these methods don't recognize that the periphery is a valid community.  MMin has a misclassification rate near 40\% in the undirected case and 30\% in the directed case. That means the model change as the increase of the shape parameter $\alpha$ doesn't affect the dividing mechanism of MMin itself here. 

\begin{figure}[t]
\centering
\begin{minipage}[t]{0.48\textwidth}
\centering
\includegraphics[width=6cm, height = 5cm]{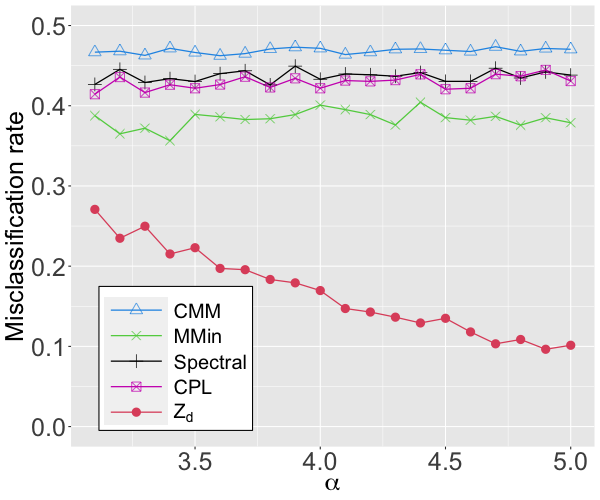}
\end{minipage}
\begin{minipage}[t]{0.48\textwidth}
\centering
\includegraphics[width=6cm, height = 5cm]{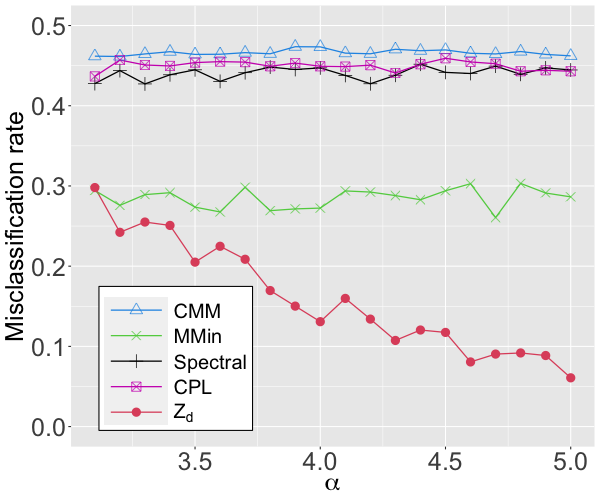}
\end{minipage}
\caption{Misclassification rate versus shape parameter of the  pareto distribution  for $N=100$ and 2 equal-sized communities. $\mathbf{P} =[0.5,0.3;0.3,0.1] $.  Left panel: undirected graph. Right panel: directed graph.}
\label{Figure:equal size,pareto,dense, core-periphery} 
\end{figure}

We have similar observations when community sizes are unbalanced, as shown in Figure \ref{Figure:unequal size,pareto,dense}. The sizes of the two communities are 30 and 70, respectively. The left panel stands for the assortative mixing case and the right panel is the core-periphery structure with the small densely inter-connected core. For the assortative mixing in the left panel, all methods are able to achieve high accuracy except for MMin, but $Z_{w}^{max}$  always achieves the  highest power in this setting. For the core-periphery case, only $Z_{d}$ can reveal the underlying community structure as expected.
\begin{figure}[htbp]
\centering
\begin{minipage}[t]{0.48\textwidth}
\centering
\includegraphics[width=6cm, height = 5cm]{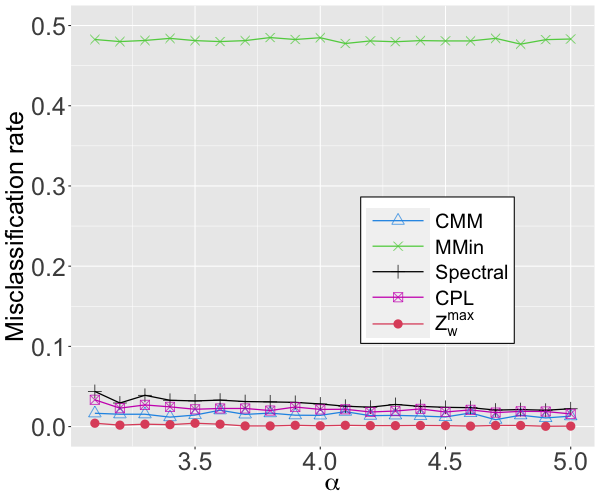}
\end{minipage}
\begin{minipage}[t]{0.48\textwidth}
\centering
\includegraphics[width=6cm, height = 5cm]{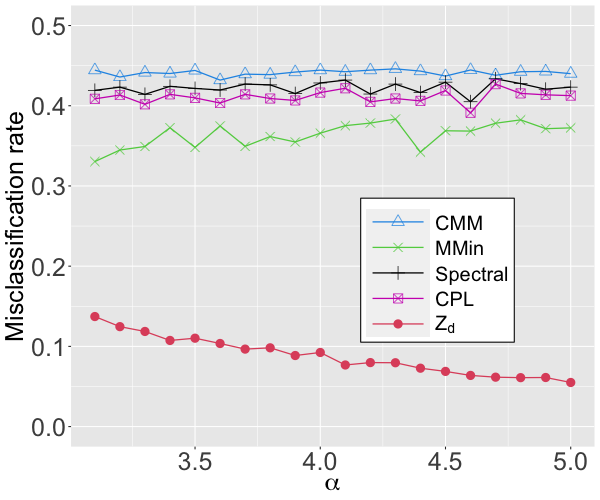}
\end{minipage}
\caption{Misclassification rate versus shape parameter of pareto $\theta$ with $m=30$ and $n=70$.   Left panel: $\mathbf{P}=[0.5,0.25;0.25,0.5]$, directed graph. Right panel: $\mathbf{P}=[0.5,0.3;0.3,0.1]$, undirected graph.}.
\label{Figure:unequal size,pareto,dense} 
\end{figure}

\begin{figure}[t]
\centering
\begin{minipage}[t]{0.48\textwidth}
\centering
\includegraphics[width=6cm,height=5cm]{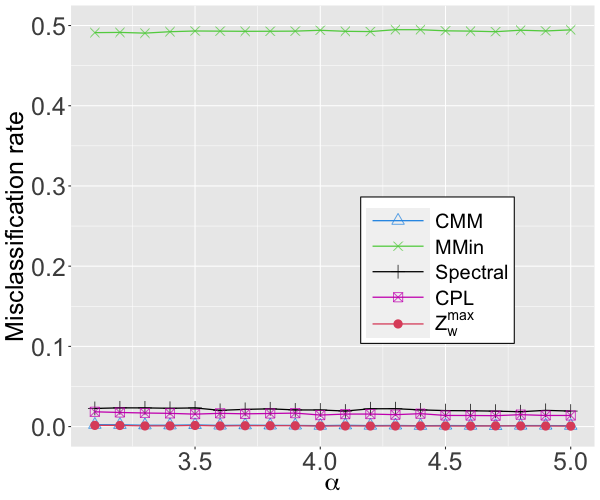}
\end{minipage}
\begin{minipage}[t]{0.48\textwidth}
\centering
\includegraphics[width=6cm, height = 5cm]{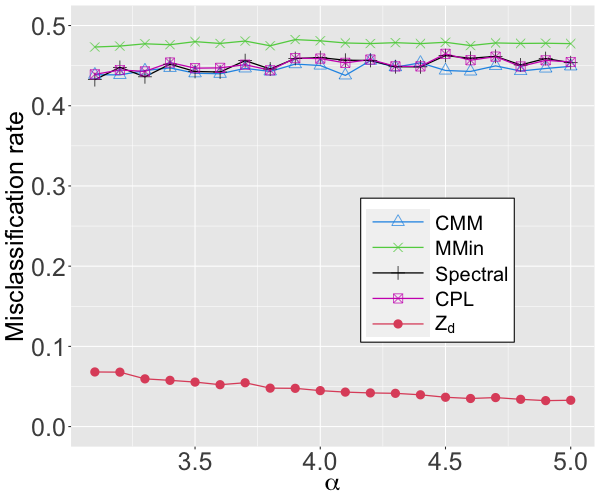}
\end{minipage}
\caption{Misclassification rate versus shape parameter of pareto $\mathbf{\theta}$.  Left panel: $\mathbf{P}=[0.1,0.05;0.05,0.1]$, $m=300, n=300$, directed graph. Right panel: $\mathbf{P}=[0.1,0.03;0.03,0.01]$, $m =100, n=500$, undirected graph. }
\label{sparse_pareto} 
\end{figure}

The performance of community detection algorithms can be affected by the density of the graph. The previous examples are all about relatively dense graphs. Here we also consider less dense graphs. In Figure \ref{sparse_pareto}, the left panel is an assortative mixing case with the connectivity matrix being $\mathbf{P} = [0.1,0.05;0.05,0.1]$ and the community sizes being $m = n = 300$. The right panel is a core-periphery structure with the connectivity matrix $\mathbf{P} = [0.1,0.03;0.03,0.01]$ and the community sizes being $m = 100, n = 500$. In Figure \ref{sparse_pareto}, left panel shows that CMM has similar power with $Z_{w}^{max}$. CPL and Spectral again suffer from the directed graph performance gap in sparse case. While for the core-periphery structure in the right panel, things remain unchanged. The statistic $Z_{d}$ is still the only one that can achieve stable high power. All the other methods are as poor as random guess.

%To sum up, $Z_{w}^{max}, Z_{w}^{min},Z_{d}$ dominate all other methods in most of the synthetic settings here including assortative mixing, disassortative mixing, core-periphery structure, directed graphs and undirected graphs, balanced communities, unbalanced communities, dense graphs and sparse graphs. 

\section{UBSea community detection framework}
\label{sec:W-D framework section}
All the simulation results in Section \ref{sec:Numerical result section} provide strong evidence that $Z_{w}^{max}, Z_{w}^{min},Z_{d}$ are  powerful when the underlying generation scheme of the  true data  lies within their consistency regions. From Theorem \ref{thm2.3}, \ref{thm_weak_consistency}, \ref{thm_strong_consistency}, it is reasonable to say that, when $N \gamma_{N}^{2} \rightarrow \infty$, we should use $Z_{d}$ as the community detection criterion; when $N\tau_{N}^{2} \rightarrow \infty$ we should use $Z_{w}^{\max}$ or $Z_{w}^{\min}$ as the criterion depend on the sign of $P_{11}+P_{22}-P_{12}-P_{21}$.  The question remains, how to choose among $Z_{w}^{max},Z_{w}^{min},Z_{d}$ if there is no prior information on the connectivity matrix available to us.  Here we discuss two possible solutions to automatically choose among these three statistics,  $\gamma$-$\tau$  criterion in Section \ref{gamma-tau section}, penalized likelihood  criterion in Section \ref{penalized likelihood section}, and check their performance in Section \ref{performance of gamma-tau and penalized likelihood}.
\subsection{\texorpdfstring{$\gamma$}{gamma}-\texorpdfstring{$\tau$}{tau} criterion}
\label{gamma-tau section}
Based on Theorem \ref{thm_weak_consistency}, when $Z_{d}$ should be used, it is expected to have a large value of $\gamma^{2}$ when we have moderate to large sample size; when $Z_{w}$ should be used, it is expected to have a large value of $\tau^{2}$. Thus, we first fit the network using $Z_{d}$, $ Z_{w}^{\max}$ and $ Z_{w}^{\min}$ separately and then calculate corresponding $\hat{\gamma}^{2}$ and $\hat{\tau}_{\max}^{2}$, $\hat{\tau}_{\min}^{2}$ values, pick the largest one as the proper statistic. We will see that this criterion, based on the consistency theory on the stochastic block model, is well-behaved when the true underlying model is exactly SBM, and to a certain extent for the DCSBM. 

\subsection{Penalized likelihood criterion}
\label{penalized likelihood section}
Likelihood method is arguable consistent under DCSBM when $\theta_{i}$'s are discrete \citep{zhao2012consistency}. However, we notice that, for the finite sample with continuous $\theta_{i}$'s case, there is always a cost associated with fitting the extra complexity of the degree-corrected model. To be more specific, if we view the heterogeneity of $\theta_{i}$'s as part of the model complexity under corrected model, the statistic $Z_{w}$ and $Z_{d}$ always try to fit a more complex model when it is not suitable for the true underlying connectivity matrix as we discussed before. Thus, inspired by the penalized likelihood raised in \cite{wang2017likelihood} for estimating the number of communities using penalized likelihood, we propose the following criterion to choose between $Z_{w}$ and $Z_{d}$:
\begin{equation}
\begin{aligned}
\label{penalized likelihood definition}
 \mathrm{L}(\mathbf{A};\boldsymbol{x}|\mathbf{P})=&\sum_{i=1}^{N}\sum_{j=1}^{N}(A_{ij}\log(P_{c_{i}(\boldsymbol{x})c_{j}(\boldsymbol{x})}\theta_{i}\theta_{j})+(1-A_{ij})\log(1-P_{c_{i}(\boldsymbol{x})c_{j}(\boldsymbol{x})}\theta_{i}\theta_{j}))\\
 &-\lambda \left(\Var(\mathbf{\theta}_{c_{1}}(\boldsymbol{x}))+\Var(\mathbf{\theta}_{c_{2}}(\boldsymbol{x}))\right)\sum_{i,j}A_{ij},
\end{aligned}
\end{equation}
where $\lambda$ is a tuning parameter, $c_{i}(\boldsymbol{x}) =1$ if $\boldsymbol{x}_{i} = 1$ and $c_{i}(\boldsymbol{x}) = 2$ if $\boldsymbol{x}_{i} = 0$,
$\mathbf{\theta}_{c_{1}}(\boldsymbol{x})$ and $\mathbf{\theta}_{c_{2}}(\boldsymbol{x})$ represent the node parameters for community 1 and community 2 determined by the label assignment $\boldsymbol{x}$, with a slight abuse of notation. Criterion (\ref{penalized likelihood definition}) is typically defined for directed graphs. For undirected graphs, we use  the same criterion with the natural undirected graph constraint $j \leq i$ in the sums.

When in use, we found that $Z_{d}$ tends to fit highly biased communities assignments when it is not the correct criterion, resulting in unexpected large value of the penalty term. To avoid the penalty term dominating the penalized  likelihood value fluctuation, and  make the criterion more stable, we replace the penalty term for $Z_{d}$ to be:
\[
\max \left(\Var(\mathbf{\theta}_{c_{1}}(\boldsymbol{x}))\left(\sum_{c_{i} = c_{j} = 1}A_{ij}\right),\Var(\mathbf{\theta}_{c_{2}}(\boldsymbol{x}))\left(\sum_{c_{i} = c_{j} = 2}A_{ij}\right)\right).
\]
Our current choice for the tuning parameter $\lambda$ is 0.12 for both directed and undirected graphs. Further investigation may be performed to analyze the choice of the tuning parameter $\lambda$.

To have an analytical expression for the MLE of $\theta_{i}$'s, we follow the trick in \cite{karrer2011stochastic}, use a poisson approximation instead of the original binomial distribution for $A_{ij}$. As a result, the MLE of $\theta_{i}$ under undirected graph is 
\begin{equation}
    \hat{\theta}_{i}=\frac{k_{i}}{\kappa_{c_{i}}},
\end{equation}
where $k_{i}$ is the degree of node $i$, $\kappa_{r}$ is the average degree in group $r$.
Similarly, the MLE of $\theta_{i}$ for directed graph has the following form:
\begin{equation}
    \hat{\theta}_{i}=\frac{k_{i}^{in} + k_{i}^{out}}{\kappa_{c_{i}}^{in} + \kappa_{c_{i}}^{out}},
\end{equation}
where $k_{i}^{in}$ is the in degree of node $i$ and $k_{i}^{out}$ is the out degree of node $i$, $\kappa_{r}^{in}$ is the average in degree of nodes in group $r$ and $\kappa_{r}^{out}$ is the average out degree of nodes in group $r$.

It is worth pointing out that for a long time period, people tend to develop the modularity-based community detection framework assuming the prior knowledge that the mixing pattern is actually assortative mixing. In essence, this implies that they prioritize maximizing modularity to identify communities, assuming that the mixing pattern is primarily assortative and  test their algorithms on assortative mixing dataset. However, researchers from different domains frequently apply modularity to disassortative or core-periphery communities and feel confused about the poor results. Therefore, it is needed to develop a framework that spends more effort discussing the different mixing types instead of simply fit a pre-determined community structure, which can benefit people from different domains with various research needs.

By adopting  the $\gamma$-$\tau$ criterion  or penalized likelihood criterion, we bring in the new insight on network community structure: There are different possible community structures (assortative mixing, disassortative mixing and core-periphery structure) for a given network simultaneously, but some are more significant.  Our task in community detection is actually first extract all the possible community structures in the current category of mixing patterns. After that, we analyze among these possible patterns, which one is more significant under some scientific rules, for instance, the $\gamma$-$\tau$ and penalized likelihood criterion in our UBSea framework.
\subsection{Performance of \texorpdfstring{$\gamma$}{gamma}-\texorpdfstring{$\tau$}{tau} criterion and penalized likelihood criterion}
\label{performance of gamma-tau and penalized likelihood}
It is relatively easier to decide whether to use $Z_{w}^{\max}$, $Z_{w}^{\min}$ or $Z_{d}$ under the Stochastic Block Model comparing to its generalization Degree Corrected Stochastic Block Model. 
\begin{figure}[b]
\centering
\begin{minipage}[t]{0.32\textwidth}
\centering
\includegraphics[scale = 0.15]{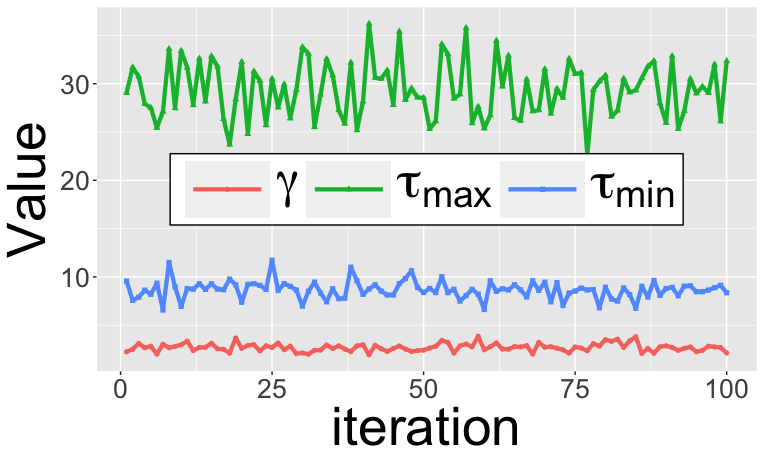}
\end{minipage}
\begin{minipage}[t]{0.32\textwidth}
\centering
\includegraphics[scale = 0.15]{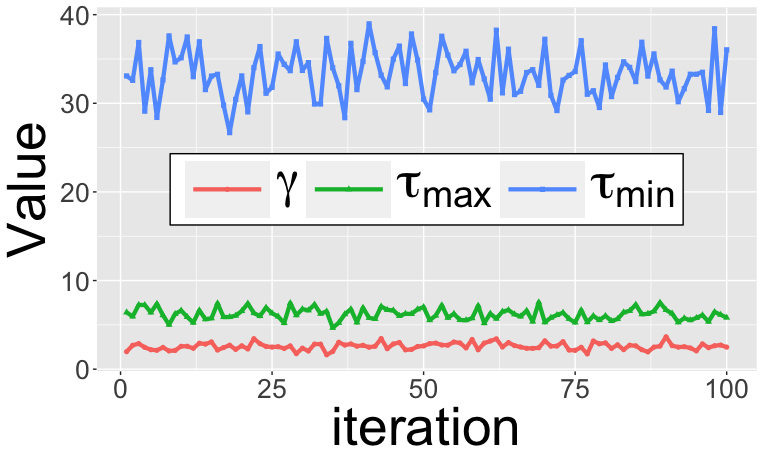}
\end{minipage}
\begin{minipage}[t]{0.32\textwidth}
\centering
\includegraphics[scale = 0.15]{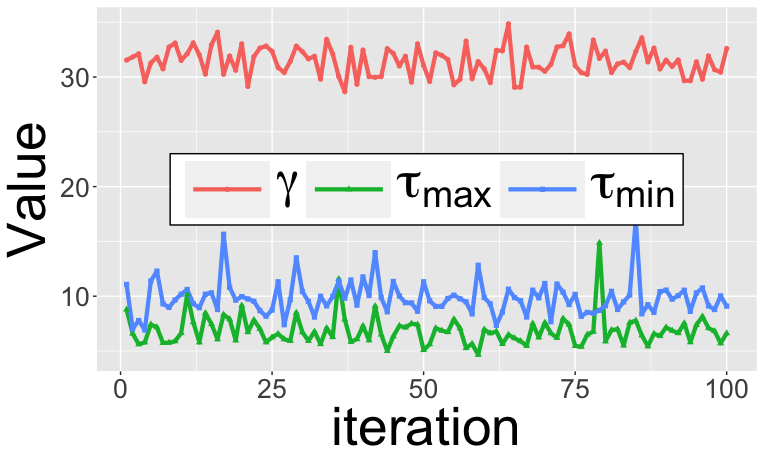}
\end{minipage}
\caption{$N \gamma^{2}$, $N\tau_{\max}^{2}$ and $N\tau_{\min}^{2}$ for $N=100$ and 2 equal-sized clusters generated from SBM model as directed graphs. From left to right: $\mathbf{P} = [0.5,0.3;0.3,0.5],\mathbf{P} = [0.3,0.5;0.5,0.3],\mathbf{P} = [0.5,0.3;0.3,0.1]$}.
\label{gamma-tau} 
\end{figure}

Figure \ref{gamma-tau} displays the corresponding values of $N\gamma^{2}$ , $N\tau_{\min}^{2}$  and $N\tau_{\max}^{2}$ for different mixing types under standard SBM. From left to right, the underlying true mixing types are  assortative mixing, disassortative mixing and core-periphery, respectively. We can see that the $\gamma$-$\tau$ criterion is pretty stable and perfect under different settings for the SBM since the values of the suitable mixing type is always the highest and much larger than the other two. For example, when the connectivity matrix $\mathbf{P} = [0.5,0.3;0.3,0.5]$ in  the left most panel, the value of $N\tau_{max}^{2}$ is significantly larger than the value of $N\tau_{min}^{2}$ and $N\gamma^{2}$. Similar results were seen in the other two panels.

However, for the more flexible DCSBM, $\gamma$-$\tau$ criterion often loses power. To  assess the performance of the $\gamma$-$\tau$ criterion under the DCSBM, we test  under different  connectivity matrices:
(I): $\mathbf{P} = 
\begin{bmatrix} 
0.1 & 0.02 \\
0.02 & 0.1
\end{bmatrix}$,
(II): $\mathbf{P} = 
\begin{bmatrix} 
0.3 & 0.5 \\
0.5 & 0.3
\end{bmatrix}$,
(III): $\mathbf{P} = 
\begin{bmatrix} 
0.5 & 0.2 \\
0.2 & 0.5
\end{bmatrix}$,
(IV): $\mathbf{P} = 
\begin{bmatrix} 
0.5 & 0.3 \\
0.3 & 0.5
\end{bmatrix}$,
(V): $\mathbf{P} = 
\begin{bmatrix} 
0.6 & 0.3 \\
0.3 & 0.1
\end{bmatrix}$.
Suppose the misclassification rate for $Z_{d}$,  $Z_{w}^{\min}$ and $ Z_{w}^{\max}$ are $\epsilon_{d}, \epsilon_{w}^{\min}$ and $\epsilon_{w}^{\max}$, respectively. Also, denote $\epsilon$ to be the misclustering rate of $\gamma$-$\tau$ criterion or penalized likelihood criterion. The success rate of the  criterion out of $s$ independent  runs is defined as:
\[
S = \frac{1}{s}\sum_{i=1}^{s}\mathbb{I}(\epsilon \leq (1 +\psi) \min\{\epsilon_{d},\epsilon_{w}^{\min}, \epsilon_{w}^{\max}\}).
\]
The tolerance rate $\psi$  is chosen to be 0.1 in the simulations below for both $\gamma$-$\tau$ criterion and penalized likelihood criterion.

\begin{figure}[b]
\centering
\begin{minipage}[t]{0.48\textwidth}
\centering
\includegraphics[width=6cm,height=5cm]{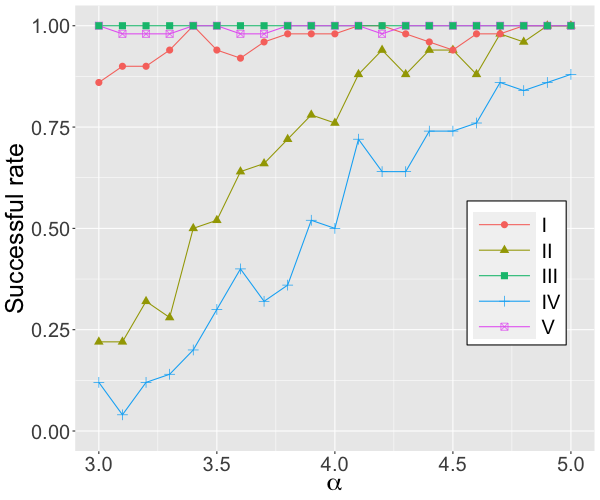}
\end{minipage}
\begin{minipage}[t]{0.48\textwidth}
\centering
\includegraphics[width=6cm, height = 5cm]{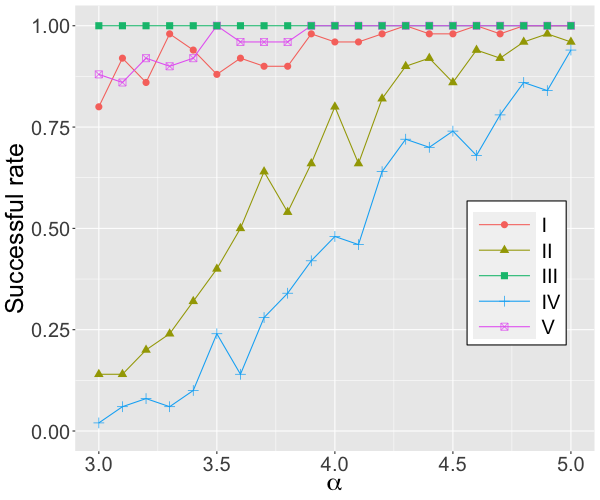}
\end{minipage}
\caption{Success rate of the $\gamma$-$\tau$ criterion, tolerance rate $\psi = 0.1$ for different settings of the connectivity matrix. The community sizes are 50 each. Left panel: undirected graphs. Right panel: directed graphs. }
\label{successful rate gamma tau} 
\end{figure}

Figure \ref{successful rate gamma tau} show that Setting IV (assortative mixing) and Setting II (disassortative mixing) are with the lowest success rate for both directed and undirected graphs, whereas Setting V (core-periphery) have relatively high success rate. Setting III is doing well because it has stronger assortative structure comparing to Setting IV. That indicates the $\gamma$-$\tau$ criterion is too aggressive in a sense that it has a preference for core-periphery structure rather than the assortative mixing pattern or disassortative mixing pattern. We want our framework, on the contrary, to have a preference for assortative mixing and disassortative mixing, and only leans toward core-periphery structure when the evidence is much more convincing. In another word, we want our framework to be more conservative for determining a core-periphery structure.  We found that penalized likelihood criterion would offer better and satisfying power under DCSBM, hence also desired under SBM if computationally fast. As a consequence, Penalized Likelihood Criterion is our main criterion for choosing between $Z_{w}$ and $Z_{d}$.  
\begin{figure}[htbp]
\centering
\begin{minipage}[t]{0.48\textwidth}
\centering
\includegraphics[width=6cm,height=5cm]{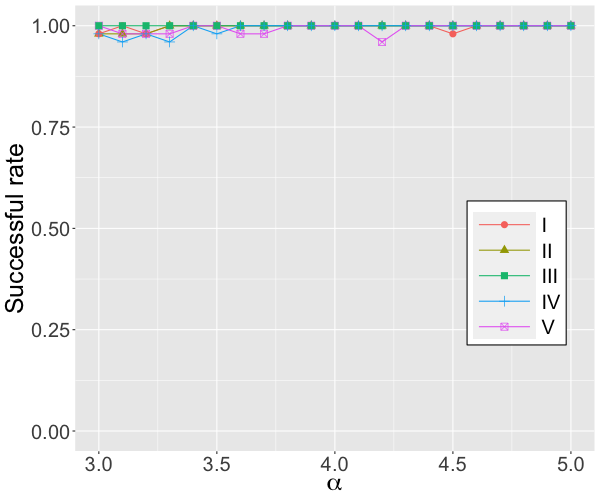}
\end{minipage}
\begin{minipage}[t]{0.48\textwidth}
\centering
\includegraphics[width=6cm, height = 5cm]{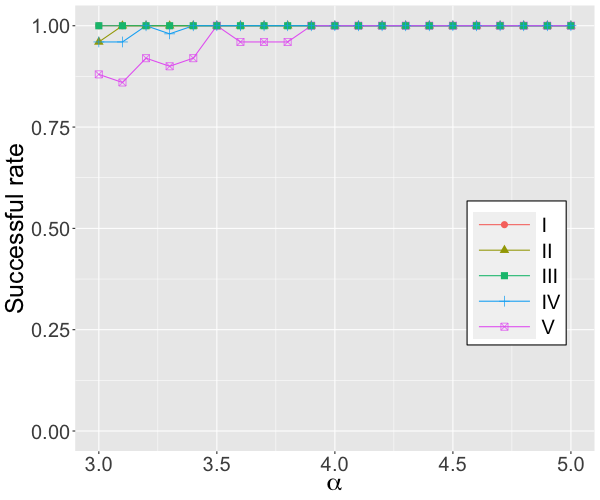}
\end{minipage}
\caption{Success rate of the penalized likelihood criterion with tuning parameter $\lambda = 0.12$ and tolerance rate $\psi = 0.1$ for different settings of the connectivity matrix. The community sizes are 50 each. Left panel: undirected graphs. Right panel: directed graphs.}
\label{successful rate} 
\end{figure}

\begin{figure}[b]
\centering
\begin{minipage}[t]{0.48\textwidth}
\centering
\includegraphics[width=6cm,height=5cm]{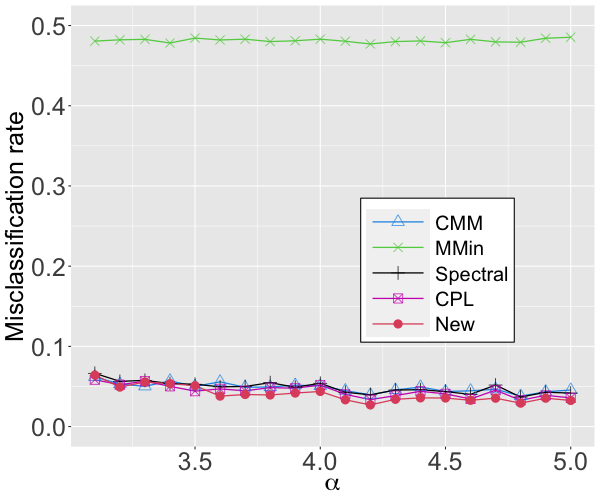}
\end{minipage}
\begin{minipage}[t]{0.48\textwidth}
\centering
\includegraphics[width=6cm, height = 5cm]{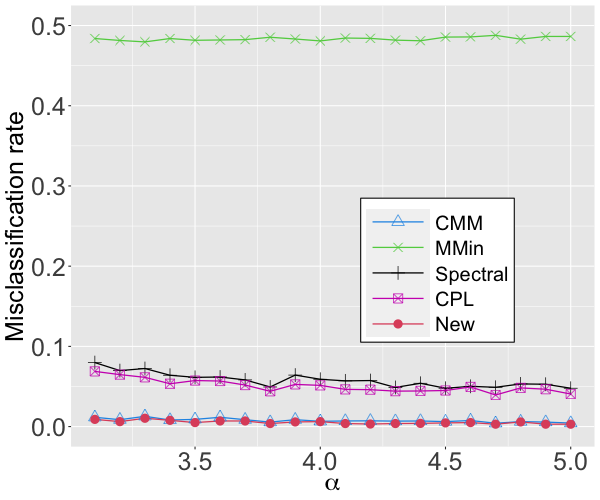}
\end{minipage}
\caption{Misclassification rate versus shape parameter of pareto $\theta$ for $N=100$ and 2 equal-sized clusters. $\mathbf{P}=[0.5,0.3;0.3,0.5]$.  Left panel: undirected graph. Right panel:directed graph.}
\label{UBSea_result_assortative} 
\end{figure}

\begin{figure}[b]
\centering
\begin{minipage}[t]{0.48\textwidth}
\centering
\includegraphics[width=6cm,height=5cm]{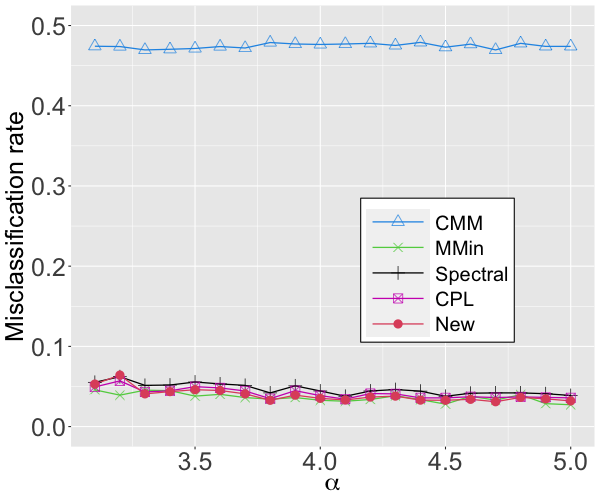}
\end{minipage}
\begin{minipage}[t]{0.48\textwidth}
\centering
\includegraphics[width=6cm, height = 5cm]{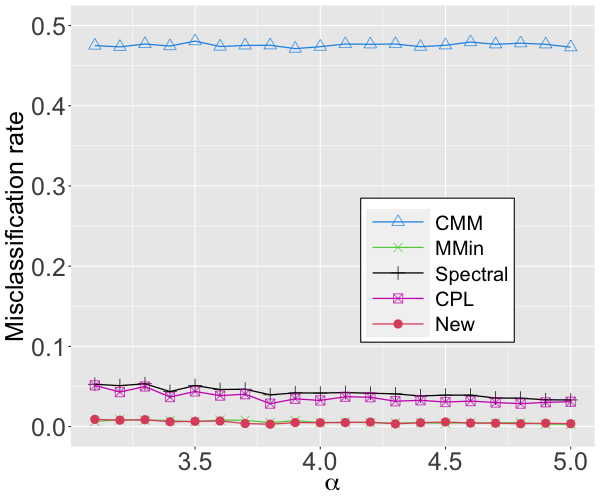}
\end{minipage}
\caption{Misclassification rate versus shape parameter of pareto $\theta$ for $N=100$ and 2 equal-sized clusters. $\mathbf{P}=[0.3,0.5;0.5,0.3]$.  Left panel: undirected graph. Right panel:directed graph.}
\label{UBSea_result_disassortative} 
\end{figure}

\begin{figure}[t]
\centering
\begin{minipage}[t]{0.48\textwidth}
\centering
\includegraphics[width=6cm,height=5cm]{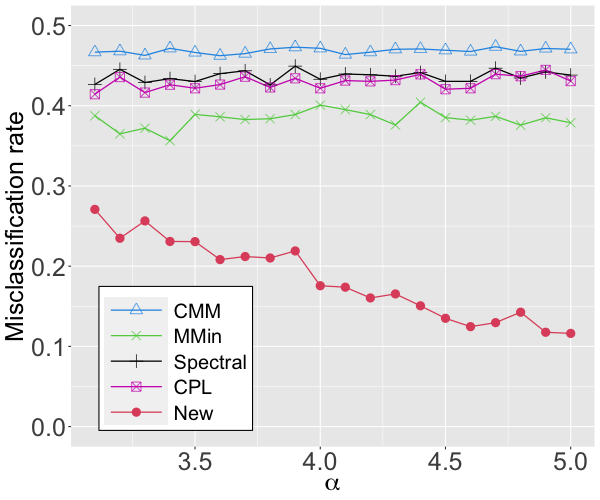}
\end{minipage}
\begin{minipage}[t]{0.48\textwidth}
\centering
\includegraphics[width=6cm, height = 5cm]{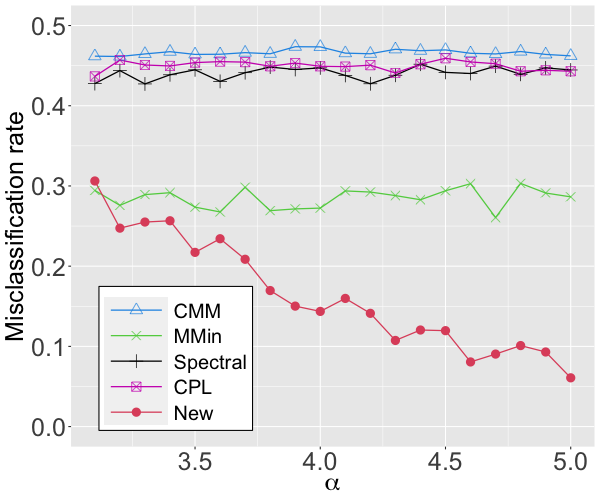}
\end{minipage}
\caption{Misclassification rate versus shape parameter of pareto $\theta$ for $N=100$ and 2 equal-sized clusters. $\mathbf{P}=[0.5,0.3;0.3,0.1]$.  Left panel: undirected graph. Right panel:directed graph.}
\label{UBSea_result_core-periphery} 
\end{figure}

Figure \ref{successful rate} show that no matter for undirected graphs or directed graphs, the success rate of penalized likelihood criterion is nearly 100\% for undirected graphs and exceeds 80\% for directed graphs. As a comparison, the $\gamma$-$\tau$ criterion attained only a power near 20\% for Setting (II) and Setting (IV) in Figure \ref{successful rate gamma tau} in the beginning of the same parameter settings. The largest value of $\theta$ when $\alpha = 3$ is often larger than 4, which means the degree heterogeneity inside every block is still considerable. Hence, the Penalized Likelihood criterion can have high power for a wide range of networks generated from  DCSBM.

Figure \ref{UBSea_result_assortative}, \ref{UBSea_result_disassortative}, \ref{UBSea_result_core-periphery} directly show the  results of our UBSea method applied to the assortative setting, disassortative setting and core-periphery for undirected graphs and directed graphs using the penalized likelihood criterion. Comparing to Figure \ref{Figure:equal size,pareto,dense, assorataive mixing}, \ref{Figure:equal size,pareto,dense, disassortative mixing}, \ref{Figure:equal size,pareto,dense, core-periphery}, the misclassification rate increased a little bit in the beginning but not much.

From now on, our default procedure for UBSea community detection will be as follows: we first fit the networks with $Z_{d},Z_{w}^{max},Z_{w}^{min}$, respectively, and then choose among the three community detection results using the penalized likelihood criterion. 

\section{Real Data Analysis}
\label{sec:real data}

\subsection{US politics book and UK faculty network data sets}
In this section, we investigate the performance of our UBSea framework on two traditional benchmarks:
\begin{itemize}
    \item the US politics book network
    \item  the UK faculty network
\end{itemize} 
 The US politics book network data was first compiled by Valdis Krebs and is unpublished. The data set can be found on Kreb's website. Nodes represents book sold on Amazon.com about US politics. Edges represent frequent co-purchasing of books by the same buyer. Nodes have been given value `l', `n' or `c' to indicate whether they are `liberal', `neutral' or `conservative'. These alignments were assigned separately by Mark Newman, we take that as the ground truth in this data set. Here we focus on the `liberal' and `conservative' books, resulting in a total number of 92 nodes and 374 undirected edges. The UK faculty network in \cite{nepusz2008fuzzy} contains a network  in a given UK university of four separate schools. The nodes represent members of the academic staff. An edge between two members represents that they are friends, according to a questionnaire. There are 33 nodes from school `1', 27 nodes from school `2', 19 nodes from school `3' and 2 nodes from school `4'. We focus on the first three school and analyze all the three combinations between two schools. For example, the sub-graph formed by the staffs from school `1' and school `2' will be analyzed as UKfaculty12 in Table \ref{Assortative real datasets}.

\begin{table}[h]
    \caption{Error rates on the Polbooks and UKfaculty data sets.}
    \label{Assortative real datasets}
    \centering
    \begin{tabular}{l|c|c|c|c}
    \hline
         Methods & Polbooks & UKfaculty12 & UKfaculty23 & UKfaculty13 \\
         \hline
         New                 & $\textbf{2/92}$   & $\textbf{1/60}$  & $\textbf{0/46}$ & \textbf{1/52}
         \\
         \hline
         CMM              & $\textbf{2/92}$   & 5/60  & 6/46 &11/52
         \\
         \hline
         CPL              & $\textbf{2/92}$   & 3/60  & 2/46 & 2/52
         \\
         \hline
         Spectral        & 3/92   & 2/60  & 1/46 & \textbf{1/52}
         \\
         \hline
         MMin              & 43/92 & 28/60 & 23/46 & 23/52
         \\
         \hline
    \end{tabular}
\end{table}

From Table \ref{Assortative real datasets}, we can see that most methods have similar behavior on these traditional benchmarks as expected, while MMin is certainly not suitable for these  traditional assortative benchmarks. For the Polbooks data set, UBSea, CMM and CPL all achieve a very low error rate, which is 2 out of 92 nodes. Spectral method performs slightly worse with one more node mis-classified. For the UK faculty data set, CMM seems to be less powerful compared with other methods. UBSea, CPL and Spectral all have high stability and accuracy on this data set.

It is worth pointing out that these benchmarks are known to be assortative mixing beforehand. Hence all the methods target on assortative mixing are expected to perform well on these benchmarks without going into discussing the mixing type step. Our UBSea framework, however, gave these community detection results based on the Penalized Likelihood criterion. We are doing this because our framework of community detection are based on a different perspective of communities mixing dynamic. We are more convinced that for a given network, assortative mixing, disassortative mixing and core-periphery structure will exist simultaneously. For these political books and UK faculty network analyzed here, the assortative mixing pattern is more significant than core-periphery structure and disassortative mixing structure as the result suggested by our UBSea framework. This is an important perspective because for instance, it is also natural to think the UK faculty network, or the general friendship networks, will exhibit the core-periphery structure with some popular people form the core and others form the periphery. Our UBSea framework confirms that  the assortative mixing evidence is much more significant than core-periphery structure on this UK faculty data set indicating by the penalized likelihood criterion.
\subsection{Word Adjacency Network}
\begin{figure}[b]
\vspace{-0.5cm}
    \centering
    \includegraphics[scale =0.25]{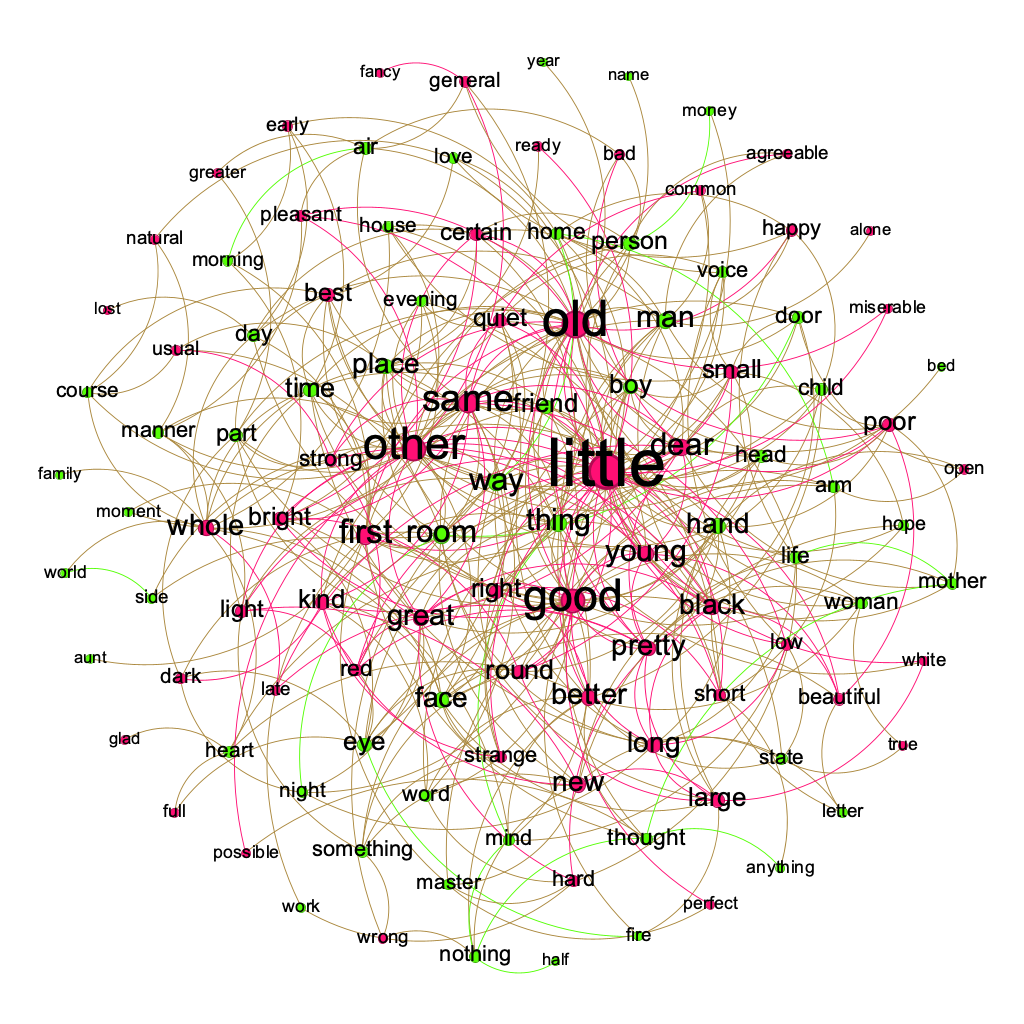}
    \vspace{-00.5cm}
    \caption{Adjective-Noun network. Red group is adjective words while green group is noun words.  Node size is scaled based on node degree. }
    \label{adj_plot}
\end{figure}This data set was compiled by \cite{newman2006finding} from the novel David Copperfield by Charles Dickens. This network consists of the 60 most commonly occurring nouns and the 60 most commonly occurring adjectives connects any pair of words that appear adjacently in the text, resulting in an undirected graph.  Excluding eight words which are disconnected from the rest leaves a network with 112 nodes. We treat `noun' and `adjective' words as two different communities and the goal is to recover these two community labels through the network structure.  The data is plotted in Figure \ref{adj_plot}.
\begin{table}[b]
    \caption{Mis-clustering rate on word-adjacency Network}
    \label{word-adjecency mis-rate}
    \centering
    \begin{tabular}{l|c|c|c|c|c}
    \hline
       Method  & New & CPL & CMM & Spectral & MMin \\
       \hline
       mis. rate & 19/112 & \textbf{18/112} &55/112&42/112 & 20/112 \\
        \hline
        \end{tabular}
\end{table}

\begin{figure}[t]
    \centering
    \includegraphics[width=10cm,height=6cm]{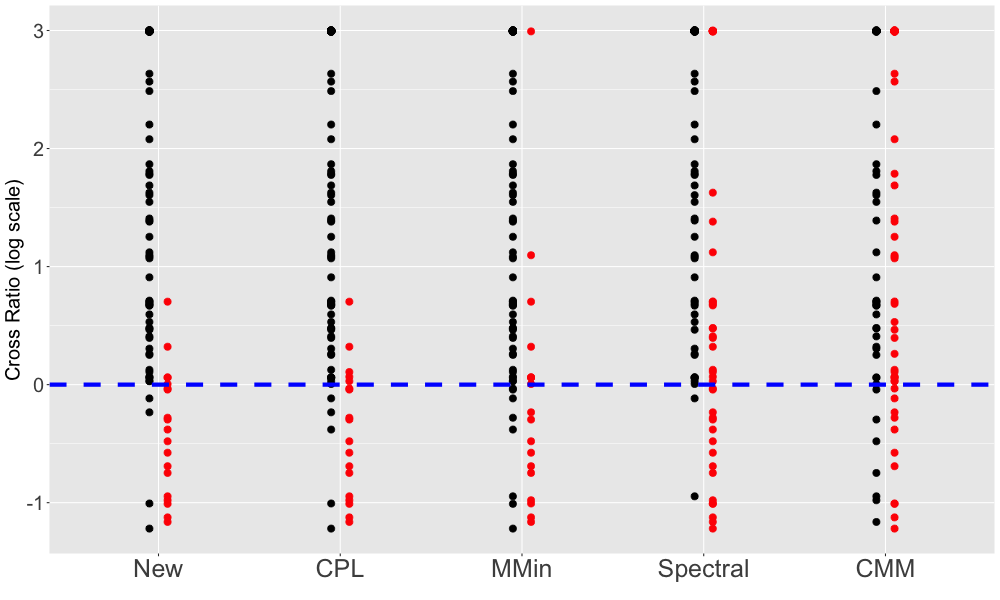}
    \caption{Scatter plots of log-scale cross ratio. Red (Black) color stands for mis-matched (matched) points comparing to the true label. Inf values are truncated to be 3.}
    \label{boxplot_disassortative}
\end{figure}

This word-adjacency network follows strong disassortative mixing pattern. It is common that an adjective word will followed by a noun. Table \ref{word-adjecency mis-rate} shows UBSea, CPL and MMin methods achieve similar accuracy rate on this data set.
%In Table \ref{comparison of UBSea,CPL and MM on word-adjacency network}, we compare the result of UBSea, CPL and MMin in detail by checking the common labels. It shows that UBSea, CPL and MMin methods agree on 41 words as ``noun'' words and 56 ``adjective'' words, which then motivate us to further investigate the remaining  disagreed words.
\iffalse
\begin{table}[htbp]
\caption{Comparison of community sizes by UBSea, CPL and Spectral methods for the word adjacency network.}
    \centering
    \begin{tabular}{l|c|c}
\hline
Method & \textbf{``Noun''} & \textbf{``Adjective''}\\
\hline
New & 49  & 63   \\ 
\hline
CPL  & 44  & 68   \\ 
\hline
MMin & 54  & 58   \\ 
\hline
New  $\cap$ CPL& 43  & 62   \\ 
\hline
New $\cap$ MMin& 47  & 56   \\ 
\hline
CPL $\cap$ MMin& 42  & 56   \\ 
\hline
New $\cap$ CPL $\cap$ MMin& 41  & 56   \\ \hline
    \end{tabular}
    \label{comparison of UBSea,CPL and MM on word-adjacency network}
\end{table}
\fi
Among these words, there are some neutral words can behave like both noun and adjective, for instance,  ``pretty, black''. To better evaluate the performance of different methods, we here define the log-scale cross ratio of node $i$ under label assignment $\boldsymbol{x}$ to be $R(i, \boldsymbol{x}) = \log \frac{b(i)}{w(i)}$, where $b(i)$ represents the number of between edges, that is, edges connect $i$ with vertices from different communities, $w(i)$ is the number of within same group edges. Figure \ref{boxplot_disassortative} shows the scatter plots of the log-scale cross ratio for different methods using the true label. Since some points don't have within community edges, i.e., $w(i) = 0$, we manually set those log-scale cross ratio to be 3 only to better visualize the results. The red points stand for the mis-matched points for different methods comparing to the underlying  truth.  We see that most of the mis-matched points for the New method have the log cross ratio less than 0, meaning that the cross edges are less than the within group edges. In this sense, these words break the disassortative structure of the network. It might be better to group the words into three groups considering these neural words, which we leave for future work. CMM totally failed to recover the underlying disassortative mixing structure. Spectral method, which  achieves slightly better performance than CMM  in Table \ref{word-adjecency mis-rate}, is again much worse than UBSea and CPL. MMin is slightly better but not as good as New and CPL. %The disassortative rates for the misclustering words are actually quite low, typically both less than 50\% for all UBSea, CPL and MMin results. This indicates the misclustering words by UBSea, CPL and MMin are mostly some neutral words. In the meantime, CMM totally failed to recover the underlying disassortative mixing structure. Spectral method, which  achieves slightly better performance than CMM  in Table \ref{word-adjecency mis-rate}, is again much worse than UBSea and CPL when checking the disassortative rate of misclustering words showing in Figure  \ref{boxplot_disassortative}.

\subsection{A co-citation network}

%\begin{figure}[!htbp]
%    \centering
%    \includegraphics[scale = 0.15]{baker.png}
%    \caption{Core/periphery structure of the adjacency network in co-citation network, created by Gephi. Size of the node reflects the connection intensity. The nodes in the center are suspicious core journals y simply observe the node sizes.}
%    \label{co-citation_structure}
%\end{figure}

This dataset was provided by \cite{baker1992structural}, who studied co-citations among social work journals on a sample of 20 social network journals included in the SSCI  Guide. His data consisted of the number of citations from one journal to another journal during a 1-year period (1985-1986), here we follow \cite{borgatti2000models} to dichotomize the data. Therefore, there will only be one directed edge from Journal A to journal B if Journal B was once cited in Journal A.  In the end, this network contains 20 nodes and 107 directed edges. We assume the identified core journals in \cite{borgatti2000models}: "SSR", "CYSR", "JSWE","SCW" and "SW" as the underlying ground truth label for core cluster.
\begin{figure}[t]
  \centering
  \includegraphics[scale = 0.18]{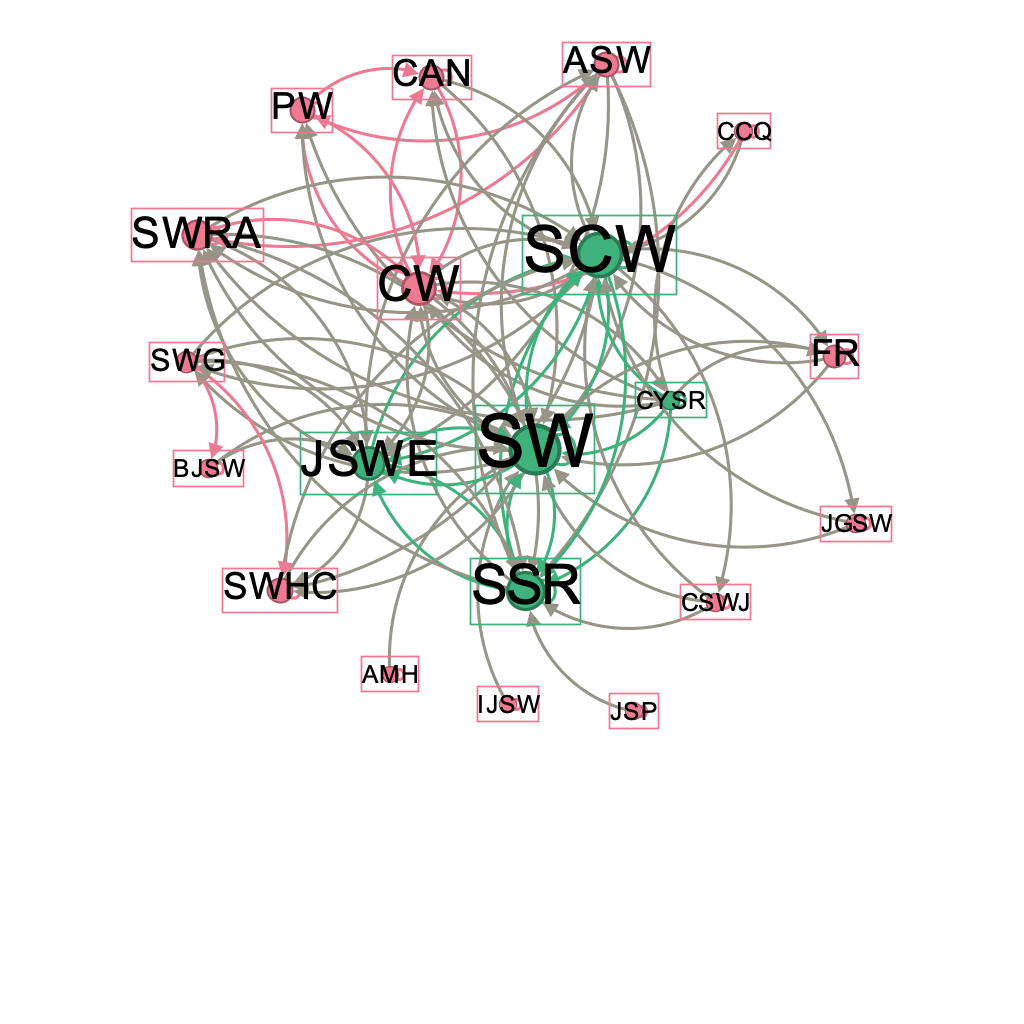}
  \vspace{-2cm}
  \caption{Visualization of co-citation network using community detection results by ``ground truth''. Red nodes  stand for the periphery and green for the core. Red edges stand for the edges inside the periphery, green edges stand for the edges inside the core and the grey ones stand for cross edges. Node size is scaled using in-degree.}
  \label{fig:co-citation1}
\end{figure}

\begin{table}[t]
    \caption{Mis-clustering rate on co-citation Network for different algorithms.mis.rate repensents the overall misclustering rate.
    mis.core stands for the misclustering rate for the core.}
    \label{mis_rate_co_citation}
    \centering
    \begin{tabular}{l|c|c|c|c|c}
    \hline
    Method  & New & CPL & CMM & Spectral & MMin\\
       \hline
       mis.rate &  \textbf{2/20} &  8/20 & 5/20 & 9/20 & 3/20 \\ 
        \hline
        mis.core & \textbf{1/5} & 4/5 &2/5& 4/5 & 2/5\\
        \hline
        \end{tabular}
\end{table}

\begin{figure}[t]
  \centering
  \includegraphics[width=0.43\textwidth]{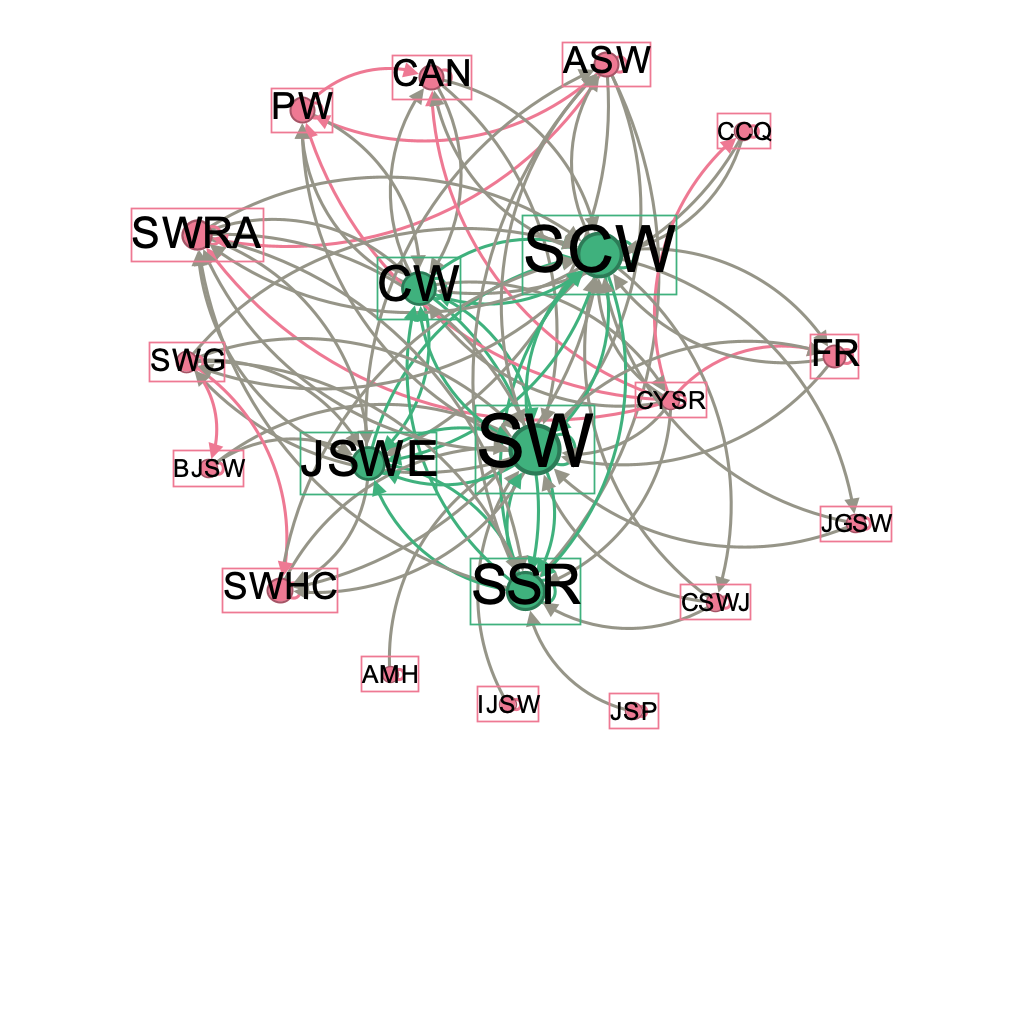}
  \hspace{1cm}
  \includegraphics[width=0.43\textwidth]{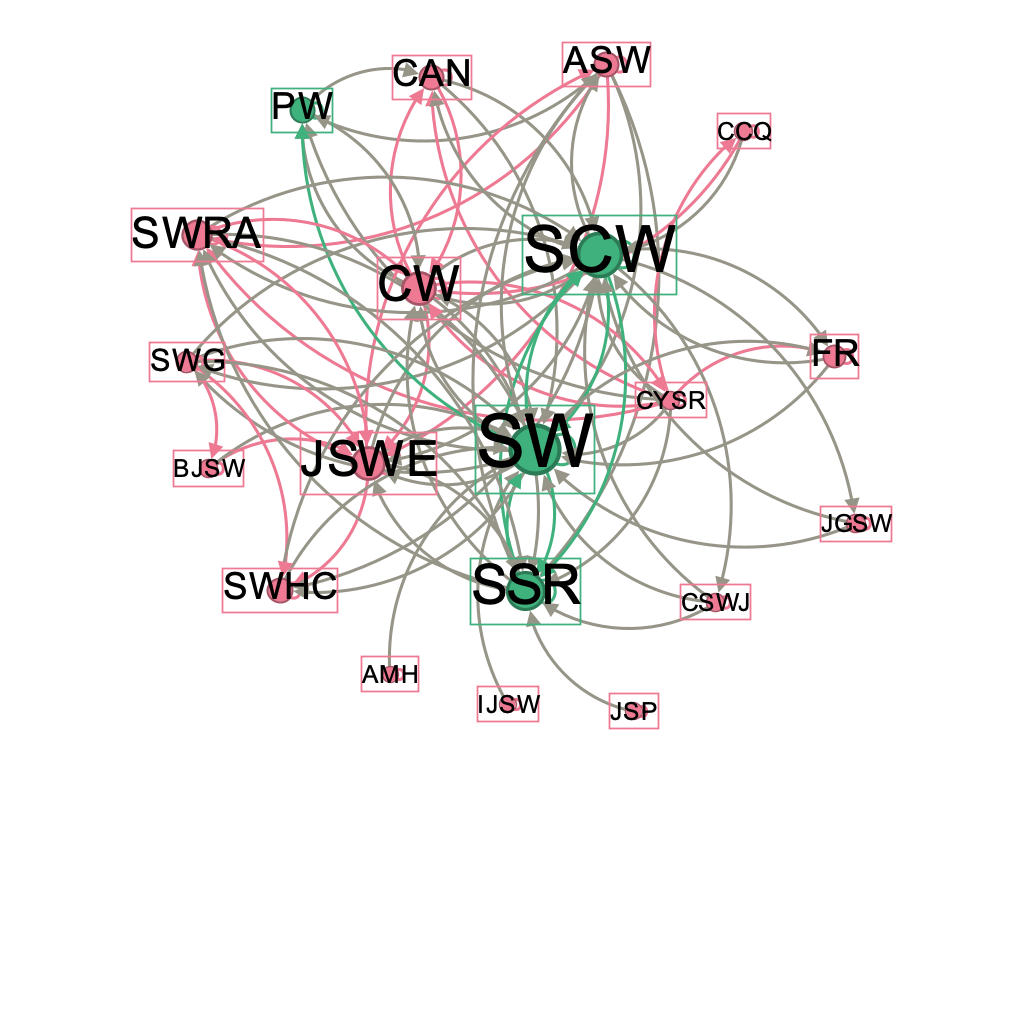}
  \vspace{-2cm}
  \caption{Visualization of co-citation network using UBSea method (left) and MMin method (right).}
  \label{fig:co-citation network, UBsea and MMin}
\end{figure}

\begin{figure}[t]
  \centering
  \includegraphics[width=0.42\textwidth]{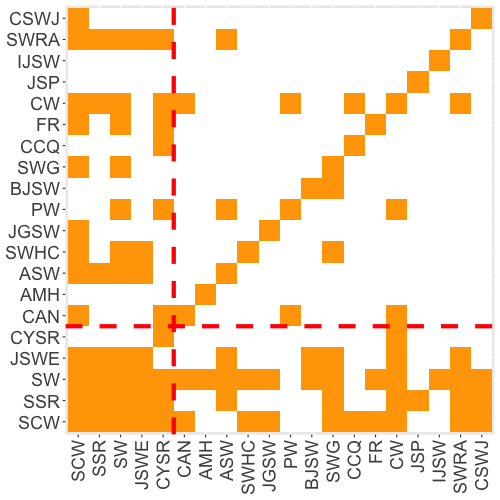}
  \hspace{1cm}
  \includegraphics[width=0.42\textwidth]{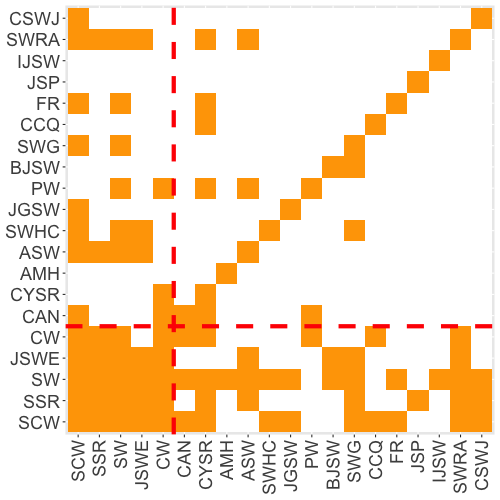}
  \caption{Left panel: adjacency plot of  the community detection result in \cite{borgatti2000models} for co-citation network.. Right panel: adjacency plot of UBSea method for co-citation network. Orange color block represents a directed edge from row journal to column journal.}
  \label{fig:co-citation_adjacency}
\end{figure}

Figure \ref{fig:co-citation1} displays the visualization for this co-citation network using community detection result in \cite{borgatti2000models}. Green edges are connections in the core, red edges are edges within the periphery and grey ones the cross edges. We see that there are very few edges (red edges) within the periphery except for one journal (``CW'') in the middle of the picture. In fact, our method puts this journal in the core instead of ``CYSR'' journal in the middle right of the picture. Table \ref{mis_rate_co_citation} displays the misclustering rate for different methods. Here, mis.rate stands for the overall misclustering rate and mis.core shows the unidentified nodes for the core found by \cite{borgatti2000models}. We see UBSea method has the lowest overall misclustering rate and can identify the most number of core nodes. CPL and Spectral produce poor results.  MMin method has fair result here. To compare the results by UBSea and MMin, Figure \ref{fig:co-citation network, UBsea and MMin} show that UBSea result is clearly better because MMin puts ``PW'' in the core and rules out both ``CW''  and ``JSWE''. It also seems more reasonable to put ``CW'' in the core by comparing Figure \ref{fig:co-citation1} and Figure \ref{fig:co-citation network, UBsea and MMin}. Figure \ref{fig:co-citation_adjacency} show the adjacency plots of ``ground truth'' and UBSea result with core in the bottom left. We see the adjacency matrix in the right panel (UBSea) has a more densely connected core.  Further investigation shows that \cite{borgatti2000models} converts the graph to  undirected  for their algorithm.

%Table \ref{mis_rate_co_citation} shows our UBSea framework algorithm can correctly uncover the underlying core-periphery structure in this network. Further investigation shows that UBSea puts CW in the core instead of CYSR, which is also reasonable based on Figure \ref{fig:co-citation1} by checking the high connectivity of CW.  CPL seems to produces low error rate but actually it only specifies one core and puts all other journals in the other community. Spectral method also loses power. CMM can perform slightly better than CPL and Spectral method. MMin is again better than CMM, Spectral and CPL in core-periphery case as in Figure \ref{UBSea_result_core-periphery}. 
  %CPL clusters 19 nodes together, results in high misclustering rate for the core. CMM and Spectral produce poor results. 

%Moreover, the co-citation network may also exhibit  assortative mixing structure in the sense that there are actually some social network journals focus on the same sub categories while others focus on other sub categories. Our UBSea framework result selected by penalized likelihood criterion suggests that the core-periphery structure is more significant in this network than assortative mixing, or even disassortative mixing structure, which gives us a better and comprehensive understanding of the network dynamic.

\section{Conclusion and Discussion}
\label{sec:conclusion and discussion}
In this paper, we studied the problem of community detection for network data with two underlying communities. We extended modularity in a strategic way and proposed a novel UBSea community detection framework. This new framework is able to detect three different kinds of mixing types: assortative mixing, disassortative mixing and core-periphery structure simultaneously. It can be applied to both directed networks and undirected networks without any generalization or extra data pre-processing steps. We provided both weak consistency and strong consistency results with similar SNR condition in the literature of SBM for the case of two communities.

After we successfully use the UBSea framework to deal with the problem of two communities. The next step is how to extend the framework to more than two communities. We develop this framework under the simple SBM with two communities. While simulation results show that this framework also enjoys extremely good performance for  data generated from the Degree-Corrected Stochastic Block Model under a wide range of settings, which can be approximated by  the stochastic block model with multiple communities, making it promising to generalize this framework to SBM with multiple communities, typically for data generated from  Binary Tree Stochastic Block Model \citep{li2022hierarchical}. We leave the task of multiple communities detection under BTSBM as part of our future work.

As a concluding remark, we want to discuss the expectation used in modularity. We here explain the reason why we want to adopt a different $\E A_{ij}$, or the null model in Section \ref{revisit modularity}. In fact, we also tried to use a modified criterion $Q_{d}$ based on modularity without changing $\E A_{ij}$, in a hope to deal with the core-periphery structure directly. Unfortunately, it does not lead to desired properties as $Z_{d}$ has. Reasons and experiments are demonstrated in Appendix \ref{appA}. Briefly speaking, in the history of community detection, the minimum cut algorithm is an early attempt by just  minimizing the cross community edges, or maximizing the edges within the communities. Modularity improves this idea by subtracting the expectation, or $\E A_{ij}$ within the communities. The expectation was calculated under an approximate configuration model so that the degree distribution under the null model is similar to the data. In another word, the degree distribution of the data becomes an additional tool in the spirit of modularity.  As we pointed out earlier, in the core-periphery case, the degree deviation measurement inside the core is positive and the deviation measurement inside the periphery is negative, hence cancelled out when added together. Thus the degree is not a useful tool in core-periphery case. Here in this paper, we adopt a much simpler null model without considering the degree distribution, but we see better performance in almost all synthetic settings and real data applications. Together with the failure of extending modularity without changing $\E A_{ij}$, it now answers the motivation for us to adopt a different $\E A_{ij}$.
%is it an additional ``tool'', or an additional ``task'', for modularity to fit the degree distribution of the data in finding the communities? 

\section{Proofs for weak and strong consistency}
\label{sec:proofs}
In this section, we prove the consistency results for $Z_{d}$ and $Z_{w}$ in Theorem \ref{thm_weak_consistency} and Theorem \ref{thm_strong_consistency}.

\theoremstyle{remark}
\begin{proof}
We first analyze $Z_{d}$ under directed graphs. Notice that
$$
\begin{aligned}
Z_{d}^{P}(\boldsymbol{x})&=\frac{R_{d}^{P}(\boldsymbol{x})-\mu_{d}^{P}(\boldsymbol{x})}{\sigma_{d}(\boldsymbol{x})}\\
&=\frac{R_{d}^{P}(\boldsymbol{x})-\mu_{d}^{P}(\boldsymbol{x})}{\sqrt{(m-\Delta_{1}+\Delta_{2})(n+\Delta_{1}-\Delta_{2})}}\sqrt{\frac{N(N-1)}{|G|+q_{1}+|G|^{2}\frac{N-4}{N}-q_{2}}}\\
&=\tilde{Z}_{d}^{P}(\boldsymbol{x})\sqrt{\frac{N(N-1)}{|G|+q_{1}+|G|^{2}\frac{N-4}{N}-q_{2}}},\\
Z_{d}(\boldsymbol{x})&=\frac{R_{d}(\boldsymbol{x})-\mu_{d}(\boldsymbol{x})}{\sigma_{d}(\boldsymbol{x})}\\
&=\frac{R_{d}(\boldsymbol{x})-\mu_{d}(\boldsymbol{x})}{\sqrt{(m-\Delta_{1}+\Delta_{2})(n+\Delta_{1}-\Delta_{2})}}\sqrt{\frac{N(N-1)}{|G|+q_{1}+|G|^{2}\frac{N-4}{N}-q_{2}}}\\
&=\tilde{Z}_{d}(\boldsymbol{x})\sqrt{\frac{N(N-1)}{|G|+q_{1}+|G|^{2}\frac{N-4}{N}-q_{2}}},\\
\end{aligned}$$
where $m,n$ are the true block sizes,  $\tilde{Z}_{d}^{P}(\boldsymbol{x}) = \frac{R_{d}^{P}(\boldsymbol{x})-\mu_{d}^{P}(\boldsymbol{x})}{\sqrt{(m-\Delta_{1}+\Delta_{2})(n+\Delta_{1}-\Delta_{2})}}$ and $\tilde{Z}_{d}(\boldsymbol{x}) = \frac{R_{d}(\boldsymbol{x})-\mu_{d}(\boldsymbol{x})}{\sqrt{(m-\Delta_{1}+\Delta_{2})(n+\Delta_{1}-\Delta_{2})}}$. Since $\sqrt{\frac{N(N-1)}{|G|+q_{1}+|G|^{2}\frac{N-4}{N}-q_{2}}}$ does not depend on $\boldsymbol{x}$, we can only consider $\Zt_{d}$ and $\tilde{Z}_{d}^{P}$ in the following context.
%With a little abuse of notation, we will write $\tilde{Z}_{d}$ as $Z_{d}$ and $Z_{w}$ correspondingly for convenience.
\begin{lemma}
If $Y_{1},\cdots,Y_{I}$ are independent, $|Y_{i}| \leq 1, \E Y_{i} = 0, S_{I} = \sum_{i=1}^{I} Y_{i}$, then
$$
    P(|S_{I}| \geq \omega) \leq 2 \exp\left(-\frac{\omega^{2}/2}{\Var(S_{I})+\omega/3}\right) \leq 2 \exp\left( - \min\left(\frac{\omega^{2}}{3\Var(S_{I})}, \frac{\omega}{2}\right)\right).
$$
\label{lemma_bernstein}
\end{lemma}
Lemma \ref{lemma_bernstein} is a special version of the well-known Bernstein concentration inequality.
Now, suppose there are two communities, with community sizes $m$ and $n$,  $N = m + n$ and the proportions $\pi_{1} = \frac{m}{N}, \pi_{2} = \frac{n}{N}$. We prove the consistency results via three steps.  Details to obtain (\ref{equ: Zd,step1}), (\ref{equ: Zd, step 2}), (\ref{equation delta}) are provided in Supplementary Material \citep{Lin2023Supplement}.\\
Step I:  Let $\epsilon_{N}$ be in the range
$$
\frac{\max(P) }{N}\ll \epsilon_{N}^{2}  \ll (2\pi_{1}P_{11}- 2\pi_{2}P_{22} - (\pi_{1}-\pi_{2})(P_{12}+P_{21}))^{2}.
$$
Notice that $\epsilon_{N}$ is well defined because the ratio between the right hand side and left hand side is $N\gamma_{N}^{2} \rightarrow \infty$. 
We prove that:
\begin{equation}
\label{equ: Zd,step1}
  P\left (\max_{\boldsymbol{x}}\frac{1}{N} |\Zt_{d}(\boldsymbol{x}) - \tilde{Z}_{d}^{P}(\boldsymbol{x})| \geq \epsilon_{N}\right ) \rightarrow 0.
\end{equation}
On one hand, if we have $\frac{N\gamma_{N}^{2}}{\log(N)} \rightarrow \infty$, we can achieve (\ref{equ: Zd,step1}) without any assumption. On the other hand, when we only have $N\gamma_{N}^{2} \rightarrow \infty$, we rule out the deviance in boundaries by assumption (\ref{assumption_1}) to achieve (\ref{equ: Zd,step1}).\\
Step II: we  prove that there exists $\delta_{N} \rightarrow 0 $ such that, when $\norm{\boldsymbol{x}-\boldsymbol{x}^{*}} \geq \delta_{N}$, we have: 
\begin{equation}
\label{equ: Zd, step 2}
   \frac{1}{N}(\tilde{Z}_{d}^{P}(\boldsymbol{x}^{*}) - \tilde{Z}_{d}^{P}(\boldsymbol{x}) )> 2\epsilon_{N},
\end{equation}
where $\norm{\boldsymbol{x}-\boldsymbol{x}^{*}}$ is defined to be  $\min\left(\frac{\Delta_{1}}{m} + \frac{\Delta_{2}}{n},  2- \frac{\Delta_{1}}{m} - \frac{\Delta_{2}}{n} \right)$. For convenience, assume $\frac{\Delta_{1}}{m} + \frac{\Delta_{2}}{n} \leq 1$. Otherwise we can  do labels permutation. Therefore, $\norm{\boldsymbol{x}-\boldsymbol{x}^{*}} =\frac{\Delta_{1}}{m} + \frac{\Delta_{2}}{n} $.\\
Furthermore, we show that  $\delta_{N}$ essentially should satisfy:
\begin{equation}
    \delta_{N} >\frac{\max (4\pi_{1},4\pi_{2})}{\sqrt{\pi_{1}\pi_{2}}} \frac{2\epsilon_{N}}{\left|2\pi_{1}P_{11}-2\pi_{2}P_{22}-(\pi_{1}-\pi_{2})(P_{12}+P_{21})\right|}.
    \label{equation delta}
\end{equation}
By the choice of $\epsilon_{N}$ in Step I, the right hand side goes to 0, thus we  enforce $\delta_{N} \rightarrow 0$.\\
Step III: Since (\ref{equ: Zd, step 2})
$$
\begin{aligned}
& P(\max_{\norm{\boldsymbol{x}-\boldsymbol{x}^{*}} > \delta_{N}}\Zt_{d}(\boldsymbol{x}) < \Zt_{d}(\boldsymbol{x}^{*}))\\
& \geq
 \quad P\left(\frac{1}{N}\max_{\norm{\boldsymbol{x}-\boldsymbol{x}^{*}} > \delta_{N}}|\Zt_{d}(\boldsymbol{x})-\tilde{Z}_{d}^{P}(\boldsymbol{x}))|< \epsilon_{N}, \frac{1}{N}|\Zt_{d}(\boldsymbol{x}^{*})-\tilde{Z}_{d}^{P}(\boldsymbol{x}^{*})| < \epsilon_{N}\right)\\ 
 &\rightarrow
 \quad 1 \quad \text{(By step I and step II),}
\end{aligned}
$$
We have the weak consistency.
\textbf{To prove strong consistency,} we need to further analyze the property within the interval $\norm{\boldsymbol{x} - \boldsymbol{x}^{*}} \leq \delta_{N}$. Without confusion, suppose that  $\frac{\Delta_{1} }{m}+ \frac{\Delta_{2}}{n} = \rho_{N}$ where $\rho_{N} \leq \delta_{N}$, $\delta_{N}$ is our specified upper bound in the previous weak consistency section. With
\[
\Zt_{d}(\boldsymbol{x}) - \Zt_{d}(\boldsymbol{x}^{*}) = (\Zt_{d}(\boldsymbol{x}) - \tilde{Z}_{d}^{P}(\boldsymbol{x})) - (\Zt_{d}(\boldsymbol{x}^{*}) - \tilde{Z}_{d}^{P}(\boldsymbol{x}^{*})) - (\tilde{Z}_{d}^{P}(\boldsymbol{x}^{*}) - \tilde{Z}_{d}^{P}(\boldsymbol{x})).
\]
Define $H = |2\pi_{1}P_{11} -2\pi_{2}P_{22}-(\pi_{1}-\pi_{2})(P_{12}+P_{21})|$. By  intermediate result in Step II,
$$
    \frac{1}{N} (\tilde{Z}_{d}^{P}(\boldsymbol{x}^{*}) - \tilde{Z}_{d}^{P}(\boldsymbol{x})) > c(\pi_{2},\pi_{2}) H \rho_{N},
$$
where $c(\pi_{1},\pi_{2})$ is some constant of $\pi_{1},\pi_{2}$. By resolving the constants in $\rho_{N}$, first we estimate the probability  that for $\norm{\boldsymbol{x}-\boldsymbol{x}^{*}} = \rho_{N}$,
\[
P\left(\left\{(\Zt_{d}(\boldsymbol{x}) - \tilde{Z}_{d}^{P}(\boldsymbol{x})) - (\Zt_{d}(\boldsymbol{x}^{*}) - \tilde{Z}_{d}^{P}(\boldsymbol{x}^{*}))\right\}  > NH\rho_{N}\right). 
\]
Case 1: if $\Delta_{1} = \Delta_{2} = \Delta$. Let
\[
A_{1} = \{(i,j)|\boldsymbol{x}_{i} = \boldsymbol{x}_{j} = 1 \},A_{0} = \{(i,j)|\boldsymbol{x}_{i} = \boldsymbol{x}_{j} = 0 \},
\]
\[
B_{1} = \{(i,j)| \boldsymbol{x}^{*}_{i} = \boldsymbol{x}^{*}_{j} = 1\},B_{0} = \{(i,j)| \boldsymbol{x}^{*}_{i} = \boldsymbol{x}^{*}_{j} = 0\}.
\]
Hence
$$
\begin{aligned}
  &(\Zt_{d}(\boldsymbol{x}) - \tilde{Z}_{d}^{P}(\boldsymbol{x})) - (\Zt_{d}(\boldsymbol{x}^{*}) - \tilde{Z}_{d}^{P}(\boldsymbol{x}^{*}))\\
=&\frac{1}{\sqrt{mn}}\left(\sum_{i,j \in (A_{1}\setminus B_{1}) \cup (B_{1}\setminus A_{1})}(A_{ij} - \E_{2} (A_{ij})) - \sum_{i,j \in (A_{0}\setminus B_{0}) \cup (B_{0}\setminus A_{0})}(A_{ij}-\E_{2} (A_{ij}))\right)  = I_{0}.
\end{aligned}
$$
By Lemma \ref{lemma_bernstein} again, 
$$P\left(|I_{0}| \geq NH \rho_{N}\right)
\lesssim \exp\left(-\min\left(t_{1}\sqrt{mn} NH \rho_{N}, t_{2}\frac{  N^{2}H^{2} \rho_{N}^{2}mn}{\max(P)(N-\Delta)\Delta} \right)\right),$$
where $t_{1},t_{2}$ are some constants. Simple calculation verified that when $||\boldsymbol{x}-\boldsymbol{x}^{*}|| < \delta_{N}$, where $\delta_{N}$ small, the second term will be the dominant term. Hence we proceed considering only the second term. Since $\frac{\Delta}{m}+\frac{\Delta}{n} = \rho_{N}$, we have that $\Delta = \frac{mn}{N}\rho_{N}\leq N\rho_{N}$.Thus,
\begin{equation}
\label{eq:inequality I1}
    P\left(\left|I_{0}\right| \geq NH \rho_{N}\right) \lesssim \exp\left(-\gamma_{N}^{2}\frac{N^{2}\rho_{N}^{2}mn}{N\rho_{N}N}\right) = \exp(-N\gamma_{N}^{2}\Delta).
\end{equation}
Case 2: If $\Delta_{1} \neq \Delta_{2}$, then
\begin{flalign*}
  &(\Zt_{d}(\boldsymbol{x}) - \tilde{Z}_{d}^{P}(\boldsymbol{x})) - (\Zt_{d}(\boldsymbol{x}^{*}) - \tilde{Z}_{d}^{P}(\boldsymbol{x}^{*})) &\\
&=\frac{1}{\sqrt{m_{\boldsymbol{x}}n_{\boldsymbol{x}}}}\left(\sum_{i,j \in A_{1}\setminus B_{1}}(A_{ij} - \E_{2}(A_{ij})) - \sum_{i,j \in A_{0} \setminus B_{0}}(A_{ij}-\E_{2}(A_{ij}))\right)\\
& \quad - \frac{1}{\sqrt{mn}}\left(\sum_{i,j \in B_{1}\setminus A_{1}}(A_{ij} - \E_{2}(A_{ij})) - \sum_{i,j \in B_{0}\setminus A_{0}}(A_{ij}-\E_{2}(A_{ij}))\right)\\
& \quad + \left(\frac{1}{\sqrt{m_{\boldsymbol{x}}n_{\boldsymbol{x}}}} - \frac{1}{\sqrt{mn}}\right)\left(\sum_{i,j \in B_{1}\cap A_{1}}(A_{ij} - \E_{2} (A_{ij})) - \sum_{i,j \in B_{0}\cap A_{0}}(A_{ij}-\E_{2} (A_{ij}))\right)\\
& = I_{1} - I_{2} + I_{3}.
\end{flalign*}
We directly analyze $I_{3}$. Since  $\frac{1}{\sqrt{m_{\boldsymbol{x}}n_{\boldsymbol{x}}}} - \frac{1}{\sqrt{mn}} = \frac{(\Delta_{2}-\Delta_{1})(m-n+\Delta_{2}-\Delta_{1})}{\sqrt{mnm_{\boldsymbol{x}}n_{\boldsymbol{x}}}(\sqrt{mn} + \sqrt{m_{\boldsymbol{x}}n_{\boldsymbol{x}}})}$ and notice that  $(1-\rho_{N})^{2}\leq m_{\boldsymbol{x}}n_{\boldsymbol{x}}/mn \leq (1+\rho_{N})^{2}$. By Lemma \ref{lemma_bernstein} again, 
\begin{flalign*}
&P\left(\left|I_{3}\right| \geq NH \rho_{N}\right)
\lesssim \exp\left(-\min\left(\frac{NH \rho_{N}(\sqrt{mn} + \sqrt{m_{\boldsymbol{x}}n_{\boldsymbol{x}}})\sqrt{mnm_{\boldsymbol{x}}n_{\boldsymbol{x}}}}{(\Delta_{2}-\Delta_{1})(m-n+\Delta_{2}-\Delta_{1})},\right. \right.&\\  
& \left.\left. \qquad \qquad \qquad \qquad \qquad \qquad \qquad 
\frac{ H^{2} \rho_{N}^{2}mnm_{\boldsymbol{x}}n_{\boldsymbol{x}}(\sqrt{mn} + \sqrt{m_{\boldsymbol{x}}n_{\boldsymbol{x}}})^{2}}{\max(P)(\Delta_{2}-\Delta_{1})^{2}(m-n+\Delta_{2}-\Delta_{1})^{2}} \right)\right).
\end{flalign*}
Plug in $\Delta_{2} -\Delta_{1} \leq \rho_{N}n \leq \rho_{N}N$ and ignore some constants, we get
$$
    P\left(\left|I_{3}\right| \geq NH \rho_{N}\right) \lesssim \exp(-N^{2}\min(H,\gamma_{N}^{2}))
$$
Furthermore, $\gamma_{N}^{2} = H \frac{H}{\max(P)}\leq 2H$, thus we have that,
\begin{equation}
\label{eq: inequality I3}
    P\left(\left|I_{3}\right| \geq NH \rho_{N}\right) \lesssim \exp(-N^{2}\min(H,\gamma_{N}^{2})) \leq \exp\left(-\frac{1}{2}N^{2}\gamma_{N}^{2}\right).
\end{equation}
$I_{1}$ and $I_{2}$ can be controlled with similar terms in (\ref{eq:inequality I1}). Combine (\ref{eq:inequality I1}) and (\ref{eq: inequality I3}), apply union bound and ignore some constants,
\begin{flalign*}
P\left(\max_{\norm{\boldsymbol{x} - \boldsymbol{x}^{*}} \leq \delta_{N}} \Zt_{d}(\boldsymbol{x}) > \Zt_{d}(\boldsymbol{x}^{*})\right)\leq
 & \sum_{\Delta_{1},\Delta_{2}}{m \choose \Delta_{1}} {n \choose \Delta_{2}} \exp\left(-\frac{1}{2}N\gamma_{N}^{2}(\Delta_{1}+\Delta_{2})\right)\\
 \leq 
 & \sum_{\Delta_{1},\Delta_{2}}{N \choose \Delta_{1} +\Delta_{2}}\exp\left(-\frac{1}{2}N\gamma_{N}^{2}(\Delta_{1}+\Delta_{2})\right)\\
 \leq
 & \sum_{\Delta = 1}^{N\delta_{N}}\Delta \exp(\Delta - \Delta \log \Delta + \Delta \log(N) - \frac{1}{2}N\gamma_{N}^{2} \Delta )),
\end{flalign*}
where $\Delta = \Delta_{1} + \Delta_{2}$. The last inequality uses Stirling approximation provided in Supplementary Material \citep{Lin2023Supplement}. Since $\Delta \geq 1$,
$$
     P\left(\max_{\norm{\boldsymbol{x} - \boldsymbol{x}^{*}} \leq \delta_{N}} \Zt_{d}(\boldsymbol{x}) > \Zt_{d}(\boldsymbol{x}^{*})\right)
     \leq 
     \sum_{\Delta=1}^{N\delta_{N}}\Delta\left(\exp\left(1+\log(N) - \frac{1}{2}N\gamma_{N}^{2}\right)\right)^{\Delta}
$$
Notice that we have:
$$
    \lim_{N \rightarrow \infty } \sum_{\Delta =1}^{\infty} \Delta \left[\exp(1+\log(N) -\frac{1}{2}N\gamma_{N}^{2})\right]^{\Delta} = \lim_{N \rightarrow \infty }\frac{\exp(1+\log(N) -\frac{1}{2}N\gamma_{N}^{2})}{(1-\exp(1+\log(N)-\frac{1}{2}N\gamma_{N}^{2}))^{2}} = 0
$$
due to the condition that $\frac{N\gamma_{N}^{2}}{\log(N)} \rightarrow \infty $. This now concludes that 
$$
    P\left(\max_{\norm{\boldsymbol{x} - \boldsymbol{x}^{*}} \leq \delta_{N}} \Zt_{d}(\boldsymbol{x}) > \Zt_{d}(\boldsymbol{x}^{*})\right)\rightarrow 0.
$$
Together with the previous result that it will achieve weak consistency without any further assumption when the condition $\frac{N\gamma_{N}^{2}}{\log(N)} \rightarrow \infty$ hold, it now completes the proof of strong consistency of $Z_{d}$.

The proof of consistency of $Z_{w}$ will based on the consistency of $Z_{d}$. To be more specific, Step I (without re-scale factor $\frac{1}{N}$), we show that
\begin{equation}
\label{equ:Zw,step1}
    P\left (\max_{\boldsymbol{x}} |\Zt_{w}(\boldsymbol{x}) - \tilde{Z}_{w}^{P}(\boldsymbol{x})| \geq \epsilon_{N}\right ) \rightarrow 0.
\end{equation}
Step II, we show that 
\begin{equation}
    \label{equ: Zw, step2}
        \tilde{Z}_{w}^{P}(\boldsymbol{x}^{*}) - \tilde{Z}_{w}^{P}(\boldsymbol{x}) > 2\epsilon_{N} \quad \text{for} \norm{\boldsymbol{x}-\boldsymbol{x}^{*}} \geq \delta_{N} \rightarrow 0
\end{equation}
for $\delta_{N}$ satisfies that 
$$
\delta_{N} > \frac{2}{\min(\pi_{1},\pi_{2})}\frac{2\epsilon_{N}}{\left|P_{11}+P_{22}-P_{12}-P_{21}\right|}.
$$
Step III, notice that
$$
\begin{aligned}
& P(\max_{\norm{\boldsymbol{x}-\boldsymbol{x}^{*}} > \delta_{N}}\Zt_{w}(\boldsymbol{x}) < \Zt_{w}(\boldsymbol{x}^{*}))\\
&\geq
\quad P\left(\max_{\norm{\boldsymbol{x}-\boldsymbol{x}^{*}} > \delta_{N}}|\Zt_{w}(\boldsymbol{x})-\tilde{Z}_{w}^{P}(\boldsymbol{x}))|< \epsilon_{N}, |\Zt_{w}(\boldsymbol{x}^{*})-\tilde{Z}_{w}^{P}(\boldsymbol{x}^{*})| < \epsilon_{N}\right)\\ & \rightarrow
\quad  1 \quad \text{(By step I and step II).}
\end{aligned}
$$
Again, we provide details of obtaining (\ref{equ:Zw,step1}) and (\ref{equ: Zw, step2}) in Supplementary Material \citep{Lin2023Supplement}. Strong consistency of $Z_{w}$ follows similar argument directly in $Z_{d}$ case and is also put in Supplementary Material \citep{Lin2023Supplement}.

All the statements above are naturally for directed graphs. But we can also extend them to undirected graphs via similar treatments as in the proof of Theorem \ref{thm2.3}. 
\end{proof}
%%%%%%%%%%%%%%%%%%%%%%%%%%%%%%%%%%%%%%%%%%%%%%
%% Example with multiple Appendixes:        %%
%%%%%%%%%%%%%%%%%%%%%%%%%%%%%%%%%%%%%%%%%%%%%%
\clearpage
\begin{appendix}

\section{A note on modularity}\label{appA}
The original definition for modularity is (\ref{equ: modularity definition}):
\[
Q= \sum_{i,j}
 \left(A_{ij}-\frac{k_{i}k_{j}}{2m}\right) \frac{(s_{i}s_{j}+ 1)}{2}.
 \]
 
As we mentioned in Section \ref{revisit modularity}, modularity maximization is able to capture the assortative mixing due to the fact that the observed edges inside each community are larger than their expected numbers. Similar observation holds for disassortative mixing. As we pointed out earlier, in the core-periphery case, the deviation measurement inside the core is positive and the deviation measurement inside the periphery is negative, hence cancelled out when added together. Thus, a natural idea for modifying modularity to adapt for the core-periphery structure then will be instead doing the subtraction, or  the following quantity:
\[
Q_{d} = \sum_{i,j}\left(A_{ij} - \frac{k_{i}k_{j}}{2m}\right)\delta_{ij},
\]
where $\delta_{ij} = 0$ if $\boldsymbol{x}_{i}\neq \boldsymbol{x}_{j}$, $\delta_{ij} = 1$ if $\boldsymbol{x}_{i} = \boldsymbol{x}_{j} = 1$, $\delta_{ij} = -1$ if $\boldsymbol{x}_{i} =\boldsymbol{x}_{j} = 0$.

\begin{figure}[b]
    \centering
    \includegraphics[width=6cm,height=5cm]{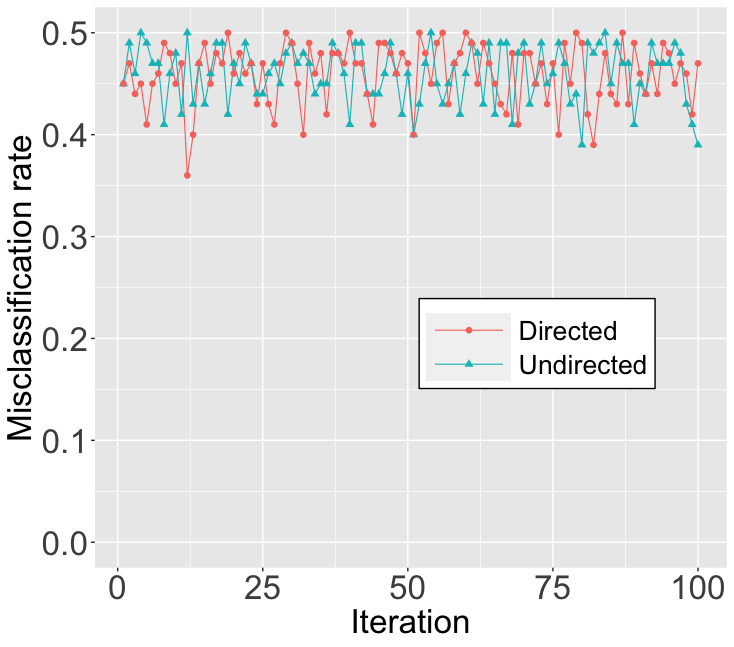}
    \caption{Misclassification rate for modified modularity under SBM with $P = [0.5,0.3;0.3,0.1]$}.
    \label{d_modularity}
\end{figure}

Figure \ref{d_modularity} shows the misclassification rate of applying $Q_{d}$ to the core-periphery example with connectivity matrix $P = [0.5,0.3;0.3,0.1]$ for both directed graphs and undirected graphs. It cleayly shows the failure of $Q_{d}$ in discovering core-periphery structure since the misclassification rate is always around 0.5.

The reasons lie in the fact that while it is true that for instance, in the core, the values of $A_{ij}$'s are larger than their expected numbers, but the corresponding  degrees $k_{i}, k_{j}$ are also relatively large. In the meantime, the values of $A_{ij}$'s are smaller than their expected numbers in the periphery, but the degrees $k_{i},k_{j}$ are also relatively small. This is different in the case of assortative and disassortative case, resulting in the insufficient large value of deviation of $Q_{d}$ at true labels assignment. Thus it is necessary for us to change the null model here or conduct other modifications in order to deal with the core-periphery structure. Our choice $Z_{d}$ is certainly a successful attempt. But it is not surprising that other choices in the category for example, the weighted difference, which is the normalized version of $R_{d} = \frac{n-1}{N-2}R_{1}(\boldsymbol{x}) - \frac{m-1}{N-2}R_{2}(\boldsymbol{x})$, exhibits higher power than $Z_{d}$ when in use since we didn't conclude $Z_{d}$ to be the optimal choice in the category of discovering the core-periphery structure.
\end{appendix}
\clearpage

\iffalse
%%%%%%%%%%%%%%%%%%%%%%%%%%%%%%%%%%%%%%%%%%%%%%
%% Support information, if any,             %%
%% should be provided in the                %%
%% Acknowledgements section.                %%
%%%%%%%%%%%%%%%%%%%%%%%%%%%%%%%%%%%%%%%%%%%%%%
\begin{acks}[Acknowledgments]
The authors would like to thank the anonymous referees, an Associate
Editor and the Editor for their constructive comments that improved the
quality of this paper.
\end{acks}

\fi
%%%%%%%%%%%%%%%%%%%%%%%%%%%%%%%%%%%%%%%%%%%%%%
%% Funding information, if any,             %%
%% should be provided in the                %%
%% funding section.                         %%
%%%%%%%%%%%%%%%%%%%%%%%%%%%%%%%%%%%%%%%%%%%%%%
\begin{funding}
The  authors were supported in part by NSF Grants DMS-1848579 and DMS-2311399.
\end{funding}

%%%%%%%%%%%%%%%%%%%%%%%%%%%%%%%%%%%%%%%%%%%%%%
%% Supplementary Material, if any, should   %%
%% be provided in {supplement} environment  %%
%% with title and short description.        %%
%%%%%%%%%%%%%%%%%%%%%%%%%%%%%%%%%%%%%%%%%%%%%%

\begin{supplement}
The supplementary materials contain proofs to theorems and lemmas, as well as additional numerical results.
\end{supplement}

%%  MR numbers will be added by VTeX.                      %%
%%                                                         %%
%%  Use \cite{...} to cite references in text.             %%
%%                                                         %%
%%%%%%%%%%%%%%%%%%%%%%%%%%%%%%%%%%%%%%%%%%%%%%%%%%%%%%%%%%%%%

%\clearpage
\bibliographystyle{imsart-nameyear}
\bibliography{ref}

\begin{thebibliography}{36}
% BibTex style file: imsart-nameyear.bst, 2017-11-03
% Default style options (sort=1,type=nameyear).
% Used options (sort=1,type=nameyear).

\bibitem[\protect\citeauthoryear{Aggarwal and Wang}{2010}]{aggarwal2010survey}
\begin{bincollection}[author]
\bauthor{\bsnm{Aggarwal},~\bfnm{Charu~C}\binits{C.~C.}} \AND
  \bauthor{\bsnm{Wang},~\bfnm{Haixun}\binits{H.}}
(\byear{2010}).
\btitle{A survey of clustering algorithms for graph data}.
In \bbooktitle{Managing and mining graph data}
\bpages{275--301}.
\bpublisher{Springer}.
\end{bincollection}
\endbibitem

\bibitem[\protect\citeauthoryear{Airoldi et~al.}{2008}]{airoldi2008mixed}
\begin{barticle}[author]
\bauthor{\bsnm{Airoldi},~\bfnm{Edo~M}\binits{E.~M.}},
  \bauthor{\bsnm{Blei},~\bfnm{David}\binits{D.}},
  \bauthor{\bsnm{Fienberg},~\bfnm{Stephen}\binits{S.}} \AND
  \bauthor{\bsnm{Xing},~\bfnm{Eric}\binits{E.}}
(\byear{2008}).
\btitle{Mixed membership stochastic blockmodels}.
\bjournal{Advances in neural information processing systems}
\bvolume{21}.
\end{barticle}
\endbibitem

\bibitem[\protect\citeauthoryear{Amini et~al.}{2013}]{amini2013pseudo}
\begin{barticle}[author]
\bauthor{\bsnm{Amini},~\bfnm{Arash~A}\binits{A.~A.}},
  \bauthor{\bsnm{Chen},~\bfnm{Aiyou}\binits{A.}},
  \bauthor{\bsnm{Bickel},~\bfnm{Peter~J}\binits{P.~J.}},
  \bauthor{\bsnm{Levina},~\bfnm{Elizaveta}\binits{E.}} \betal{et~al.}
(\byear{2013}).
\btitle{Pseudo-likelihood methods for community detection in large sparse
  networks}.
\bjournal{Annals of Statistics}
\bvolume{41}
\bpages{2097--2122}.
\end{barticle}
\endbibitem

\bibitem[\protect\citeauthoryear{Baker}{1992}]{baker1992structural}
\begin{barticle}[author]
\bauthor{\bsnm{Baker},~\bfnm{Donald~R}\binits{D.~R.}}
(\byear{1992}).
\btitle{A structural analysis of social work journal network: 1985-1986}.
\bjournal{Journal of Social Service Research}
\bvolume{15}
\bpages{153--168}.
\end{barticle}
\endbibitem

\bibitem[\protect\citeauthoryear{Bickel and
  Chen}{2009}]{bickel2009nonparametric}
\begin{barticle}[author]
\bauthor{\bsnm{Bickel},~\bfnm{Peter~J}\binits{P.~J.}} \AND
  \bauthor{\bsnm{Chen},~\bfnm{Aiyou}\binits{A.}}
(\byear{2009}).
\btitle{A nonparametric view of network models and Newman--Girvan and other
  modularities}.
\bjournal{Proceedings of the National Academy of Sciences}
\bvolume{106}
\bpages{21068--21073}.
\end{barticle}
\endbibitem

\bibitem[\protect\citeauthoryear{Borgatti and
  Everett}{2000}]{borgatti2000models}
\begin{barticle}[author]
\bauthor{\bsnm{Borgatti},~\bfnm{Stephen~P}\binits{S.~P.}} \AND
  \bauthor{\bsnm{Everett},~\bfnm{Martin~G}\binits{M.~G.}}
(\byear{2000}).
\btitle{Models of core/periphery structures}.
\bjournal{Social networks}
\bvolume{21}
\bpages{375--395}.
\end{barticle}
\endbibitem

\bibitem[\protect\citeauthoryear{Cai and Li}{2015}]{cai2015robust}
\begin{barticle}[author]
\bauthor{\bsnm{Cai},~\bfnm{T~Tony}\binits{T.~T.}} \AND
  \bauthor{\bsnm{Li},~\bfnm{Xiaodong}\binits{X.}}
(\byear{2015}).
\btitle{Robust and computationally feasible community detection in the presence
  of arbitrary outlier nodes}.
\bjournal{The Annals of Statistics}
\bvolume{43}
\bpages{1027--1059}.
\end{barticle}
\endbibitem

\bibitem[\protect\citeauthoryear{Chen, Chen and Su}{2018}]{chen2018weighted}
\begin{barticle}[author]
\bauthor{\bsnm{Chen},~\bfnm{Hao}\binits{H.}},
  \bauthor{\bsnm{Chen},~\bfnm{Xu}\binits{X.}} \AND
  \bauthor{\bsnm{Su},~\bfnm{Yi}\binits{Y.}}
(\byear{2018}).
\btitle{A weighted edge-count two-sample test for multivariate and object
  data}.
\bjournal{Journal of the American Statistical Association}
\bvolume{113}
\bpages{1146--1155}.
\end{barticle}
\endbibitem

\bibitem[\protect\citeauthoryear{Chen and Friedman}{2017}]{chen2017new}
\begin{barticle}[author]
\bauthor{\bsnm{Chen},~\bfnm{Hao}\binits{H.}} \AND
  \bauthor{\bsnm{Friedman},~\bfnm{Jerome~H}\binits{J.~H.}}
(\byear{2017}).
\btitle{A new graph-based two-sample test for multivariate and object data}.
\bjournal{Journal of the American statistical association}
\bvolume{112}
\bpages{397--409}.
\end{barticle}
\endbibitem

\bibitem[\protect\citeauthoryear{Chen, Li and Xu}{2018}]{chen2018convexified}
\begin{barticle}[author]
\bauthor{\bsnm{Chen},~\bfnm{Yudong}\binits{Y.}},
  \bauthor{\bsnm{Li},~\bfnm{Xiaodong}\binits{X.}} \AND
  \bauthor{\bsnm{Xu},~\bfnm{Jiaming}\binits{J.}}
(\byear{2018}).
\btitle{Convexified modularity maximization for degree-corrected stochastic
  block models}.
\bjournal{Annals of Statistics}
\bvolume{46}
\bpages{1573--1602}.
\end{barticle}
\endbibitem

\bibitem[\protect\citeauthoryear{Chen and Lin}{2023}]{chen2023new}
\begin{barticle}[author]
\bauthor{\bsnm{Chen},~\bfnm{Hao}\binits{H.}} \AND
  \bauthor{\bsnm{Lin},~\bfnm{Xiancheng}\binits{X.}}
(\byear{2023}).
\btitle{A new clustering framework}.
\bjournal{arXiv preprint arXiv:2305.00578}.
\end{barticle}
\endbibitem

\bibitem[\protect\citeauthoryear{Fortunato and
  Barthelemy}{2007}]{fortunato2007resolution}
\begin{barticle}[author]
\bauthor{\bsnm{Fortunato},~\bfnm{Santo}\binits{S.}} \AND
  \bauthor{\bsnm{Barthelemy},~\bfnm{Marc}\binits{M.}}
(\byear{2007}).
\btitle{Resolution limit in community detection}.
\bjournal{Proceedings of the national academy of sciences}
\bvolume{104}
\bpages{36--41}.
\end{barticle}
\endbibitem

\bibitem[\protect\citeauthoryear{Holland, Laskey and
  Leinhardt}{1983}]{holland1983stochastic}
\begin{barticle}[author]
\bauthor{\bsnm{Holland},~\bfnm{Paul~W}\binits{P.~W.}},
  \bauthor{\bsnm{Laskey},~\bfnm{Kathryn~Blackmond}\binits{K.~B.}} \AND
  \bauthor{\bsnm{Leinhardt},~\bfnm{Samuel}\binits{S.}}
(\byear{1983}).
\btitle{Stochastic blockmodels: First steps}.
\bjournal{Social networks}
\bvolume{5}
\bpages{109--137}.
\end{barticle}
\endbibitem

\bibitem[\protect\citeauthoryear{Karrer and
  Newman}{2011}]{karrer2011stochastic}
\begin{barticle}[author]
\bauthor{\bsnm{Karrer},~\bfnm{Brian}\binits{B.}} \AND
  \bauthor{\bsnm{Newman},~\bfnm{Mark~EJ}\binits{M.~E.}}
(\byear{2011}).
\btitle{Stochastic blockmodels and community structure in networks}.
\bjournal{Physical review E}
\bvolume{83}
\bpages{016107}.
\end{barticle}
\endbibitem

\bibitem[\protect\citeauthoryear{Krzakala et~al.}{2013}]{krzakala2013spectral}
\begin{barticle}[author]
\bauthor{\bsnm{Krzakala},~\bfnm{Florent}\binits{F.}},
  \bauthor{\bsnm{Moore},~\bfnm{Cristopher}\binits{C.}},
  \bauthor{\bsnm{Mossel},~\bfnm{Elchanan}\binits{E.}},
  \bauthor{\bsnm{Neeman},~\bfnm{Joe}\binits{J.}},
  \bauthor{\bsnm{Sly},~\bfnm{Allan}\binits{A.}},
  \bauthor{\bsnm{Zdeborov{\'a}},~\bfnm{Lenka}\binits{L.}} \AND
  \bauthor{\bsnm{Zhang},~\bfnm{Pan}\binits{P.}}
(\byear{2013}).
\btitle{Spectral redemption in clustering sparse networks}.
\bjournal{Proceedings of the National Academy of Sciences}
\bvolume{110}
\bpages{20935--20940}.
\end{barticle}
\endbibitem

\bibitem[\protect\citeauthoryear{Leicht and Newman}{2008}]{leicht2008community}
\begin{barticle}[author]
\bauthor{\bsnm{Leicht},~\bfnm{Elizabeth~A}\binits{E.~A.}} \AND
  \bauthor{\bsnm{Newman},~\bfnm{Mark~EJ}\binits{M.~E.}}
(\byear{2008}).
\btitle{Community structure in directed networks}.
\bjournal{Physical review letters}
\bvolume{100}
\bpages{118703}.
\end{barticle}
\endbibitem

\bibitem[\protect\citeauthoryear{Li et~al.}{2022}]{li2022hierarchical}
\begin{barticle}[author]
\bauthor{\bsnm{Li},~\bfnm{Tianxi}\binits{T.}},
  \bauthor{\bsnm{Lei},~\bfnm{Lihua}\binits{L.}},
  \bauthor{\bsnm{Bhattacharyya},~\bfnm{Sharmodeep}\binits{S.}},
  \bauthor{\bparticle{Van~den} \bsnm{Berge},~\bfnm{Koen}\binits{K.}},
  \bauthor{\bsnm{Sarkar},~\bfnm{Purnamrita}\binits{P.}},
  \bauthor{\bsnm{Bickel},~\bfnm{Peter~J}\binits{P.~J.}} \AND
  \bauthor{\bsnm{Levina},~\bfnm{Elizaveta}\binits{E.}}
(\byear{2022}).
\btitle{Hierarchical community detection by recursive partitioning}.
\bjournal{Journal of the American Statistical Association}
\bvolume{117}
\bpages{951--968}.
\end{barticle}
\endbibitem

\bibitem[\protect\citeauthoryear{Lin and Chen}{2023}]{Lin2023Supplement}
\begin{barticle}[author]
\bauthor{\bsnm{Lin},~\bfnm{Xiancheng}\binits{X.}} \AND
  \bauthor{\bsnm{Chen},~\bfnm{Hao}\binits{H.}}
(\byear{2023}).
\btitle{Supplement to ``UBSEA: A Unified Community Detection Framework''}.
\end{barticle}
\endbibitem

\bibitem[\protect\citeauthoryear{Liu and Chen}{2022}]{liu2022fast}
\begin{barticle}[author]
\bauthor{\bsnm{Liu},~\bfnm{Yi-Wei}\binits{Y.-W.}} \AND
  \bauthor{\bsnm{Chen},~\bfnm{Hao}\binits{H.}}
(\byear{2022}).
\btitle{A fast and efficient change-point detection framework based on
  approximate k-nearest neighbor graphs}.
\bjournal{IEEE Transactions on Signal Processing}.
\end{barticle}
\endbibitem

\bibitem[\protect\citeauthoryear{Malliaros and
  Vazirgiannis}{2013}]{malliaros2013clustering}
\begin{barticle}[author]
\bauthor{\bsnm{Malliaros},~\bfnm{Fragkiskos~D}\binits{F.~D.}} \AND
  \bauthor{\bsnm{Vazirgiannis},~\bfnm{Michalis}\binits{M.}}
(\byear{2013}).
\btitle{Clustering and community detection in directed networks: A survey}.
\bjournal{Physics reports}
\bvolume{533}
\bpages{95--142}.
\end{barticle}
\endbibitem

\bibitem[\protect\citeauthoryear{Moore et~al.}{2011}]{moore2011active}
\begin{binproceedings}[author]
\bauthor{\bsnm{Moore},~\bfnm{Cristopher}\binits{C.}},
  \bauthor{\bsnm{Yan},~\bfnm{Xiaoran}\binits{X.}},
  \bauthor{\bsnm{Zhu},~\bfnm{Yaojia}\binits{Y.}},
  \bauthor{\bsnm{Rouquier},~\bfnm{Jean-Baptiste}\binits{J.-B.}} \AND
  \bauthor{\bsnm{Lane},~\bfnm{Terran}\binits{T.}}
(\byear{2011}).
\btitle{Active learning for node classification in assortative and
  disassortative networks}.
In \bbooktitle{Proceedings of the 17th ACM SIGKDD international conference on
  Knowledge discovery and data mining}
\bpages{841--849}.
\end{binproceedings}
\endbibitem

\bibitem[\protect\citeauthoryear{Nepusz et~al.}{2008}]{nepusz2008fuzzy}
\begin{barticle}[author]
\bauthor{\bsnm{Nepusz},~\bfnm{Tam{\'a}s}\binits{T.}},
  \bauthor{\bsnm{Petr{\'o}czi},~\bfnm{Andrea}\binits{A.}},
  \bauthor{\bsnm{N{\'e}gyessy},~\bfnm{L{\'a}szl{\'o}}\binits{L.}} \AND
  \bauthor{\bsnm{Bazs{\'o}},~\bfnm{F{\"u}l{\"o}p}\binits{F.}}
(\byear{2008}).
\btitle{Fuzzy communities and the concept of bridgeness in complex networks}.
\bjournal{Physical Review E}
\bvolume{77}
\bpages{016107}.
\end{barticle}
\endbibitem

\bibitem[\protect\citeauthoryear{Newman}{2002}]{newman2002assortative}
\begin{barticle}[author]
\bauthor{\bsnm{Newman},~\bfnm{Mark~EJ}\binits{M.~E.}}
(\byear{2002}).
\btitle{Assortative mixing in networks}.
\bjournal{Physical review letters}
\bvolume{89}
\bpages{208701}.
\end{barticle}
\endbibitem

\bibitem[\protect\citeauthoryear{Newman}{2003a}]{newman2003mixing}
\begin{barticle}[author]
\bauthor{\bsnm{Newman},~\bfnm{Mark~EJ}\binits{M.~E.}}
(\byear{2003}a).
\btitle{Mixing patterns in networks}.
\bjournal{Physical review E}
\bvolume{67}
\bpages{026126}.
\end{barticle}
\endbibitem

\bibitem[\protect\citeauthoryear{Newman}{2003b}]{newman2003structure}
\begin{barticle}[author]
\bauthor{\bsnm{Newman},~\bfnm{Mark~EJ}\binits{M.~E.}}
(\byear{2003}b).
\btitle{The structure and function of complex networks}.
\bjournal{SIAM review}
\bvolume{45}
\bpages{167--256}.
\end{barticle}
\endbibitem

\bibitem[\protect\citeauthoryear{Newman}{2006a}]{newman2006modularity}
\begin{barticle}[author]
\bauthor{\bsnm{Newman},~\bfnm{Mark~EJ}\binits{M.~E.}}
(\byear{2006}a).
\btitle{Modularity and community structure in networks}.
\bjournal{Proceedings of the national academy of sciences}
\bvolume{103}
\bpages{8577--8582}.
\end{barticle}
\endbibitem

\bibitem[\protect\citeauthoryear{Newman}{2006b}]{newman2006finding}
\begin{barticle}[author]
\bauthor{\bsnm{Newman},~\bfnm{Mark~EJ}\binits{M.~E.}}
(\byear{2006}b).
\btitle{Finding community structure in networks using the eigenvectors of
  matrices}.
\bjournal{Physical review E}
\bvolume{74}
\bpages{036104}.
\end{barticle}
\endbibitem

\bibitem[\protect\citeauthoryear{Newman and Girvan}{2004}]{newman2004finding}
\begin{barticle}[author]
\bauthor{\bsnm{Newman},~\bfnm{Mark~EJ}\binits{M.~E.}} \AND
  \bauthor{\bsnm{Girvan},~\bfnm{Michelle}\binits{M.}}
(\byear{2004}).
\btitle{Finding and evaluating community structure in networks}.
\bjournal{Physical review E}
\bvolume{69}
\bpages{026113}.
\end{barticle}
\endbibitem

\bibitem[\protect\citeauthoryear{Newman and Leicht}{2007}]{newman2007mixture}
\begin{barticle}[author]
\bauthor{\bsnm{Newman},~\bfnm{Mark~EJ}\binits{M.~E.}} \AND
  \bauthor{\bsnm{Leicht},~\bfnm{Elizabeth~A}\binits{E.~A.}}
(\byear{2007}).
\btitle{Mixture models and exploratory analysis in networks}.
\bjournal{Proceedings of the National Academy of Sciences}
\bvolume{104}
\bpages{9564--9569}.
\end{barticle}
\endbibitem

\bibitem[\protect\citeauthoryear{Peixoto}{2018}]{peixoto2018nonparametric}
\begin{barticle}[author]
\bauthor{\bsnm{Peixoto},~\bfnm{Tiago~P}\binits{T.~P.}}
(\byear{2018}).
\btitle{Nonparametric weighted stochastic block models}.
\bjournal{Physical Review E}
\bvolume{97}
\bpages{012306}.
\end{barticle}
\endbibitem

\bibitem[\protect\citeauthoryear{Qin and Rohe}{2013}]{qin2013regularized}
\begin{barticle}[author]
\bauthor{\bsnm{Qin},~\bfnm{Tai}\binits{T.}} \AND
  \bauthor{\bsnm{Rohe},~\bfnm{Karl}\binits{K.}}
(\byear{2013}).
\btitle{Regularized spectral clustering under the degree-corrected stochastic
  blockmodel}.
\bjournal{Advances in neural information processing systems}
\bvolume{26}.
\end{barticle}
\endbibitem

\bibitem[\protect\citeauthoryear{Reichardt and
  Bornholdt}{2006}]{reichardt2006statistical}
\begin{barticle}[author]
\bauthor{\bsnm{Reichardt},~\bfnm{J{\"o}rg}\binits{J.}} \AND
  \bauthor{\bsnm{Bornholdt},~\bfnm{Stefan}\binits{S.}}
(\byear{2006}).
\btitle{Statistical mechanics of community detection}.
\bjournal{Physical review E}
\bvolume{74}
\bpages{016110}.
\end{barticle}
\endbibitem

\bibitem[\protect\citeauthoryear{Von~Luxburg}{2007}]{von2007tutorial}
\begin{barticle}[author]
\bauthor{\bsnm{Von~Luxburg},~\bfnm{Ulrike}\binits{U.}}
(\byear{2007}).
\btitle{A tutorial on spectral clustering}.
\bjournal{Statistics and computing}
\bvolume{17}
\bpages{395--416}.
\end{barticle}
\endbibitem

\bibitem[\protect\citeauthoryear{Wang and Bickel}{2017}]{wang2017likelihood}
\begin{barticle}[author]
\bauthor{\bsnm{Wang},~\bfnm{YX~Rachel}\binits{Y.~R.}} \AND
  \bauthor{\bsnm{Bickel},~\bfnm{Peter~J}\binits{P.~J.}}
(\byear{2017}).
\btitle{Likelihood-based model selection for stochastic block models}.
\bjournal{The Annals of Statistics}
\bvolume{45}
\bpages{500--528}.
\end{barticle}
\endbibitem

\bibitem[\protect\citeauthoryear{Zhang, Martin and
  Newman}{2015}]{zhang2015identification}
\begin{barticle}[author]
\bauthor{\bsnm{Zhang},~\bfnm{Xiao}\binits{X.}},
  \bauthor{\bsnm{Martin},~\bfnm{Travis}\binits{T.}} \AND
  \bauthor{\bsnm{Newman},~\bfnm{Mark~EJ}\binits{M.~E.}}
(\byear{2015}).
\btitle{Identification of core-periphery structure in networks}.
\bjournal{Physical Review E}
\bvolume{91}
\bpages{032803}.
\end{barticle}
\endbibitem

\bibitem[\protect\citeauthoryear{Zhao, Levina and
  Zhu}{2012}]{zhao2012consistency}
\begin{barticle}[author]
\bauthor{\bsnm{Zhao},~\bfnm{Yunpeng}\binits{Y.}},
  \bauthor{\bsnm{Levina},~\bfnm{Elizaveta}\binits{E.}} \AND
  \bauthor{\bsnm{Zhu},~\bfnm{Ji}\binits{J.}}
(\byear{2012}).
\btitle{Consistency of community detection in networks under degree-corrected
  stochastic block models}.
\bjournal{The Annals of Statistics}
\bvolume{40}
\bpages{2266--2292}.
\end{barticle}
\endbibitem

\end{thebibliography}
\clearpage

\end{document}